\documentclass[10pt,onecolumn]{article}

\usepackage{adjustbox}
\usepackage{graphicx}
 
\usepackage{color}

\usepackage{amsfonts}
\usepackage{amssymb}

\usepackage{caption}
\usepackage{comment}
\usepackage{authblk}
\usepackage{pifont}
\usepackage{amsthm}
\usepackage{color, colortbl}

\usepackage{geometry}
\usepackage{hyperref}
\usepackage{cite}

\usepackage[cmex10]{amsmath}

\usepackage{url}

\newenvironment{myitemize}{
\begin{itemize}
 \setlength{\itemsep}{1pt}
 \setlength{\parskip}{0pt}
 \setlength{\parsep}{0pt}}{\end{itemize}
}

\theoremstyle{plain}
\newtheorem{theorem}{Theorem}[section]

\newtheorem{cor}[theorem]{Corollary}
\newtheorem{prop}[theorem]{Proposition}
\theoremstyle{definition}
\newtheorem{definition}{Definition}[section]

\theoremstyle{remark}

\begin{document}

\title{\vspace{-3.5cm}A Decomposition Method for Global Evaluation of Shannon Entropy and Local Estimations of Algorithmic Complexity}

\author[1,2,3]{\normalsize Hector~Zenil\footnote{Corresponding author: hector.zenil AT algorithmicnaturelab.org\\Source code: \url{https://www.algorithmicdynamics.net/software.html}}}
\author[1,4]{Santiago~Hern\'andez-Orozco}
\author[1,3]{Narsis~A.~Kiani}
\author[3,5]{Fernando~Soler-Toscano}
\author[3,6]{Antonio~Rueda-Toicen} 
\author[7,8]{Jesper Tegn\'er}

\affil[1]{Algorithmic Dynamics Lab, Unit of Computational Medicine, Department of Medicine Solna, Center for Molecular Medicine, Karolinska Institute and SciLifeLab, Stockholm, Sweden}
\affil[2]{Department of Computer Science, University of Oxford, U.K.}
\affil[3]{Algorithmic Nature Group, LABORES, Paris, France}
\affil[4]{Posgrado en Ciencia e Ingenier\'ia de la Computaci\'{o}n, Universidad Nacional Aut\'onoma de M\'exico (UNAM), Mexico City, Mexico}
\affil[5]{Grupo de L\'ogica, Lenguaje e Informaci\'on, Universidad de Sevilla, Spain}
\affil[6]{Instituto Nacional de Bioingenier\'ia, Universidad Central de Venezuela, Caracas}
\affil[7]{Unit of Computational Medicine, Department of Medicine Solna, Center for Molecular Medicine, SciLifeLab and Karolinska Institute, Stockholm, Sweden}
\affil[8]{Biological and Environmental Sciences and Engineering Division, Computer, Electrical and Mathematical Sciences and Engineering Division, King Abdullah University of Science and Technology (KAUST), Kingdom of Saudi Arabia}

\date{}

\maketitle

\vspace{-1cm}

\begin{abstract}
We investigate the properties of a Block Decomposition Method (BDM), which extends the power of a Coding Theorem Method (CTM) that approximates local estimations of algorithmic complexity based upon Solomonoff-Levin's theory of algorithmic probability providing a closer connection to algorithmic complexity than previous attempts based on statistical regularities such as popular lossless compression schemes. The strategy behind BDM is to find small computer programs that produce the components of a larger, decomposed object. The set of short computer programs can then be artfully arranged in sequence so as to produce the original object. We show that the method provides efficient estimations of algorithmic complexity but that it performs like Shannon entropy when it loses accuracy. We estimate errors and study the behaviour of BDM for different boundary conditions, all of which are compared and assessed in detail. The measure may be adapted for use with more multi-dimensional objects than strings, objects such as arrays and tensors. To test the measure we demonstrate the power of CTM on low algorithmic-randomness objects that are assigned maximal entropy (e.g. $\pi$) but whose numerical approximations are closer to the theoretical low algorithmic-randomness expectation. We also test the measure on larger objects including dual, isomorphic and cospectral graphs for which we know that algorithmic randomness is low. We also release implementations of the methods in most major programming languages---\textit{Wolfram Language} (Mathematica), \textit{Matlab}, \textit{R}, \textit{Perl}, \textit{Python}, \textit{Pascal}, \textit{C++}, and \textit{Haskell}---and an online algorithmic complexity calculator.

\end{abstract}


\section{Measuring Complexity}

Capturing the `complexity' of an object, for purposes such as classification and object profiling, is one of the most fundamental challenges in science. This is so because one has to either choose a computable measure (e.g. Shannon entropy) that is not invariant to object descriptions and probability distributions~\cite{zenilkianipre} and lacks an \textit{invariance theorem}--- which forces one to decide on a particular feature shared among several objects of interest---or else estimate values of an uncomputable function in applying a `universal' measure of complexity that is invariant to object description (such as \textit{algorithmic complexity}). 

This latter drawback has led to computable variants and the development of time- and resource-bounded algorithmic complexity/probability that is finitely computable~\cite{levin73,Daley,Daley2,Schmidhuber}. A good introduction and list of references is provided in~\cite{li}. Here we study a measure that lies half-way between two universally used measures that enables the action of both at different scales  by dividing data into smaller pieces for which the halting problem involved in an uncomputable function can be partially circumvented, in exchange for a huge calculation based upon the concept of algorithmic probability. The calculation can, however, be precomputed and hence reused in future applications, thereby constituting a strategy for efficient estimations--bounded by Shannon entropy and by algorithmic (Kolmogorov-Chaitin) complexity--in exchange for a loss of accuracy.

In the past, lossless compression algorithms have dominated the landscape of applications of algorithmic complexity. When researchers have chosen to use lossless compression algorithms for reasonably long strings, the method has proven to be of value (e.g.~\cite{rivalsmcilibrasimli}). Their successful application has to do with the fact that compression is a sufficient test for non-algorithmic randomness (though the converse is not true). However, popular implementations of lossless compression algorithms are based upon estimations of \textit{entropy}~\cite{emergence}, and are therefore no more closely related to algorithmic complexity than is Shannon entropy by itself. They can only account for statistical regularities and not for algorithmic ones, though accounting for algorithmic regularities ought to be crucial, since these regularities represent the main advantage of using algorithmic complexity. 

One of the main difficulties with computable measures of complexity such as Shannon entropy is that they are not robust enough ~\cite{zenilkianipre,zenildata}. For example, they are not invariant to different descriptions of the same object--unlike algorithmic complexity, where the so-called \textit{invariance theorem} guarantees the invariance of an object's algorithmic complexity. This is due to the fact that one can always translate a lossless description into any other lossless description simply with a program of a fixed length, hence in effect just adding a constant. Computability theorists are not much concerned with the relatively negligible differences between evaluations of Kolmogorov complexity which are owed to the use of different descriptive frameworks (e.g. different programming languages), yet these differences are fundamental in applications of algorithmic complexity. 

Here we study a Block Decomposition Method (BDM) that is meant to extend the power of the so-called \textit{Coding Theorem Method} (CTM). Applications of CTM include image classification~\cite{kolmo2d} and visual cognition~\cite{gauvrit1,gauvrit2,kempe}, among many applications in cognitive science. In these applications, other complexity measures, including entropy and lossless compressibility, have been outperformed by CTM. Graph complexity is another subject of active research~\cite{Holzinger,Dehmer,Dehmer2,Dehmer3,Dehmer4}, the method here presented has  made a contribution to this subject proposing robust measures of algorithmic graph complexity~\cite{zenilgraph,zenilmethodsbiology}.

After introducing the basics of algorithmic complexity, algorithmic probability, Shannon entropy and other entropic measures, and after exploring the limitations and abuse of the use of lossless compression to approximate algorithmic complexity, we introduce CTM on which BDM heavily relies upon. After introducing BDM, we thoroughly study its properties and parameter dependency on size, the problem of the boundaries in the decomposition process, we prove theoretical bounds and provide numerical estimations after testing on actual data (graphs) followed by error estimations for the numerical estimations by BDM. We also introduce a normalized version of BDM.

\subsection{Algorithmic Complexity}

The \textit{Coding Theorem Method} was first introduced as a method for dealing with the problem of compressing very short strings, for which no implementation of lossless compression gives reasonable results. CTM exploits the elegant and powerful relationship between the algorithmic frequency of production of a string and its algorithmic complexity~\cite{levin}.

The \emph{algorithmic complexity}~\cite{kolmo,chaitin} $K(s)$ of a string $s$ is the length of the shortest program $p$ that outputs the string $s$, when running on a universal (prefix-free \footnote{The group of valid programs forms a prefix-free set (no element is a prefix of any other, a property necessary to keep $0 < m(s) < 1$). For details see~\cite{calude,li}.}) Turing machine $U$. 

A technical inconvenience of $K$ as a function taking $s$ to be the length of the shortest program that produces $s$, is that $K$ is lower semi-computable. In other words, there is no effective algorithm which takes a string $s$ as input and produces the integer $K(s)$ as output. This is usually considered a major problem, but the theory of algorithmic randomness~\cite{downey} ascribes uncomputability to any universal measure of complexity, that is, a measure that is at least capable of characterizing mathematical randomness~\cite{martinlof}. However, because it is lower semi-computable, $K(s)$ can be approximated from above, or in other words, upper bounds can be found, for example, by finding and exhibiting a small computer program (measured in bits) relative to the length of a bit string.

\subsection{Algorithmic Probability}
\label{algop}

The classical probability of production of a bit string $s$ among all possible $2^n$ bit strings of length $n$ is given by $P(s) = 1/2^n$. The concept of algorithmic probability (also known as Levin's semi-measure) replaces the random production of outputs by the random production of programs that produce an output. The algorithmic probability of a string $s$ is thus a measure that estimates the probability of a random program $p$ producing a string $s$ when run on a universal (prefix-free) Turing machine $U$. 

The algorithmic probability $m(s)$ of a binary string $s$ is the sum over all the (prefix-free) programs $p$ for which a universal Turing machine $U$ running $p$ outputs $s$ and halts~\cite{solomonoff,levin,chaitin}. It replaces $n$ (the length of $s$) with $|p|$, the length of the program $p$ that produces $s$:

\begin{equation}
\label{m}
m(s) = \sum_{p:U(p) = s} 1/2^{|p|}
\end{equation}

$m(s)$ can be considered an approximation to $K(s)$, because the greatest contributor to $m(s)$ is the shortest program $p$ that generates $s$ using $U$. So if $s$ is of low algorithmic complexity, then $|p|<n$, and will be considered random if $|p|\sim n$.

\begin{theorem}
The Coding Theorem~\cite{solomonoff,levin} further establishes the connection between $m(s)$ and $K(s)$.
\begin{equation}\label{cdthm}
|-\log_2 m(s) - K(s)| < c
\end{equation}
\noindent where $c$ is a fixed constant, independent of $s$.
\end{theorem}

The Coding Theorem implies~\cite{cover, caludelibro} that the output frequency distribution of random computer programs to approximate $m(s)$ can be clearly converted into estimations of $K(s)$ using the following rewritten version of Eq. (~\ref{cdthm}):

\begin{equation}~\label{codingeq}
K(s)=-\log_2 m(s) + O(1)
\end{equation}

Among the properties of Algorithmic Probability and $m(s)$ that makes it optimal is that the data does not need to be stationary or ergodic and is universal (stronger than ergodic) in the sense that it will work for any string and can deal with missing and multidimensional data~\cite{solo,solo2,solo3,kirchner}, there is no underfitting or overfitting because the method is parameter free and the data need not to be divided into training and test sets.

\subsection{Convergence Rate and the \textit{Invariance Theorem}}

One other fundamental property that provides the theory of algorithmic complexity with the necessary robustness to stand as a universal measure of (random) complexity is the so-called \textit{Invariance theorem}~\cite{solomonoff,levin}, which guarantees the convergence of values despite the use of different reference universal Turing machines (UTMs) or e.g. programming languages. 

\begin{theorem}
Invariance theorem~\cite{solomonoff,kolmo,chaitin}: If $U_1$ and $U_2$ are two UTMs and $K_{U_1}(s)$ and $K_{U_2}(s)$ the algorithmic complexity of $s$ for $U_1$ and $U_2$, there exists a constant $c$ such that: 
\begin{equation}
\label{invariance}
| K_{U_1}(s) - K_{U_2}(s) | < c_{_{U_1,U_2}}
\end{equation}
\end{theorem}

\noindent where $c_{_{U_1,U_2}}$ is independent of $s$ and can be considered to be the length (in bits) of a translating function between universal Turing machines $U_1$ and $U_2$, or as a compiler between computer programming languages $U_1$ and $U_1$. 

\begin{figure}[ht!]
\begin{center}
  \makebox[\columnwidth]{\includegraphics[width=13cm]{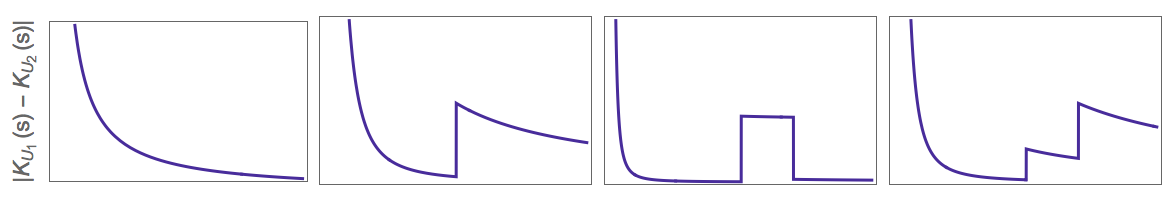}}
\end{center}
\caption{\label{convergence}Hypothetical behaviour of (non-)regular convergence rates of the constant involved in the \textit{invariance theorem}. The invariance theorem guarantees that complexity values for a string $s$ measured by different \textit{reference} UTMs $U_1$ and $U_2$ will only diverge by a constant $c$ (the length between $U_1$ and $U_2$) independent of $s$ yet it does not tell how fast or in what way the convergence may happen particularly at the beginning. The invariance theorem only tells us that at the limit the curve will converge to a small and constant value $c$, but it tells us nothing about the rate of convergence or about transitional behaviour.}
\end{figure}

In practice, however, the constant involved can be arbitrarily large, and the invariance theorem tells us nothing about the convergence (see Fig.~\ref{convergence}). One may perform the calculation of $K(s)$ for a growing sequence $s$ under $U_1$ in the expectation that for long $s$, $K_{U_1}(s) = K(s)$. However, there is no guarantee that this will be the case, and the size of $c_{_{U_1,U_2}}$ in general is unknown. 

It is a question whether there can be a \textit{natural} universal Turing machine $U_N$ such that $K_{U_N}(s)$ converges faster for $s$ than for any other universal Turing machine (UTM), or whether specific conditions must be met if $U_1$ is to generate  \textit{``well-behaved"} (monotonic) behaviour in $c$~\cite{ctm}. The \textit{invariance theorem} guarantees that such optimal `well-behaved' machines $U_N$ always exist--indeed their existence is implicit in the very sense of the theorem (meaning any universal machine can be optimal in the sense of the theorem)-- but it tells nothing about the rate of convergence or about transitional behaviour (see Fig.~\ref{convergence} for illustration).

The longer the string, the less important $c$ is (i.e. the choice of programming language or UTM). However, in practice $c$ can be arbitrarily large, thus having a great impact, particularly on short strings, and never revealing at which point one starts approaching a stable $K$ or when one is diverging before finally monotonically converging, as is seen in the different possible behaviours illustrated in the sketches in Fig.~\ref{convergence}.

The invariance theorem tells us that it is impossible to guarantee convergence but it does not imply that one cannot study the behaviour of such a constant for different reference universal Turing machines nor that $K$ cannot be approximated from above. 

\section{The Use and Misuse of Lossless Compression}

Notice that the same problem affects compression algorithms as they are widely used to approximate $K$. They are not exempt from the same constant problem. Lossless compression is also subject to the constant involved in the \textit{invariance theorem}, because there is no reason to choose one compression algorithm over another.

Lossless compression algorithms have traditionally been used to approximate the Kolmogorov complexity of an object (e.g. a string) because they can provide upper bounds to $K$ and compression is sufficient test for non-randomness. In a similar fashion, our approximations are upper bounds based on finding a small Turing machine producing a string. Data compression can be viewed as a function that maps data onto other data using the same units or alphabet (if the translation is into different units or a larger or smaller alphabet, then the process is called an encoding).

Compression is successful if the resulting data are shorter than the original data plus the decompression instructions needed to fully reconstruct said original data. For a compression algorithm to be lossless, there must be a reverse mapping from compressed data to the original data. That is to say, the compression method must encapsulate a bijection between ``plain" and ``compressed" data, because the original data and the compressed data should be in the same units. By a simple counting argument, lossless data compression algorithms cannot guarantee compression for all input data sets, because there will be some inputs that do not get smaller when processed by the compression algorithm, and for any lossless data compression algorithm that makes at least one file smaller, there will be at least one file that it makes larger. Strings of data of length $N$ or shorter are clearly a strict superset of the sequences of length $N-1$ or shorter. It follows therefore that there are more data strings of length $N$ or shorter than there are data strings of length $N - 1$ or shorter. And it follows from the \emph{pigeonhole principle} that it is not possible to map every sequence of length $N$ or shorter to a unique sequence of length $N - 1$ or shorter. Therefore there is no single algorithm that reduces the size of all data. 

One of the more time consuming steps of implementations of, for example, LZ77 compression (one of the most popular) is the search for the longest string match. Most lossless compression implementations are based upon the LZ algorithm. The classical LZ77 and LZ78 algorithms enact a greedy parsing of the input data. That is, at each step, they take the longest dictionary phrase which is a prefix of the currently unparsed string suffix. LZ algorithms are said to be `universal' because, assuming unbounded memory (arbitrary sliding window length), they asymptotically approximate the (infinite) entropy rate of the generating source~\cite{lempelziv}. Not only does lossless compression fail to provide any estimation of the algorithmic complexity of small objects~\cite{delahayezenil,ctm}, it is also not more closely related to algorithmic complexity than Shannon entropy by itself~\cite{emergence}, being only capable of exploiting statistical regularities (if the observer has no other method to update/infer the probability distribution)~\cite{zenildata}.

The greatest limitation of popular lossless compression algorithms, in the light of algorithmic complexity, is that their implementations only exploit statistical regularities (repetitions up to the size of the sliding window length). Thus in effect no general lossless compression algorithm does better than provide the Shannon entropy rate (c.f. Section~\ref{entropyrate}) of the objects it compresses. It is then obvious that an exploration of other possible methods for approximating $K$ is not only desirable but needed, especially methods that can, at least, in principle, and more crucially in practice, detect algorithmic features in data that statistical approaches such as Entropy and to some extent compression would miss.

\subsection{Building upon Block Entropy}
\label{entropyrate}

The entropy $Η$ of a discrete random variable $s$ with possible values ${s_1, \dots, s_n}$ and probability distribution $P(s)$ is defined as:

\begin{definition}
    $$H(s)=-\sum_{i=1}^n P(s_i) \log_2 P(s_i)$$
\end{definition}

\noindent In the case of $P(s_i) = 0$ for some $i$, the value of the corresponding summand 0 $log_2(0)$ is taken to be 0.

It is natural to ask how random a string appears when blocks of finite length are considered. 

For example, the string $01010101\ldots01$ is periodic, but for the smallest granularity (1 bit) or 1-symbol block, the sequence has maximal entropy, because the number of 0s and 1s is the same assuming a uniform probability distribution for all strings of the same finite length. Only for longer blocks of length 2 bits can the string be found to be regular, identifying the smallest entropy value for which the granularity is at its minimum.


When dealing with a given string $s$, assumed to originate from a stationary stochastic source with known probability density for each symbol, the following function $H_l$ gives what is variously denominated as block entropy and is Shannon entropy over blocks (or subsequences of $s$) of length $l$. That is, 

\begin{definition}
    $$H_l(s)= - \displaystyle\sum_{b \in blocks} P_l(b) \log_2 P_l(b),$$
\end{definition}
\noindent{}where $blocks$ is the set resulting of decomposing $s$ in substrings or blocks of size $l$ and $P_l(b)$ is the probability of obtaining the combination of $n$ symbols corresponding to the block $b$. For infinite strings assumed to originate from a stationary source, the \textit{entropy rate} of $s$ can be defined as the limit $$\lim_{l\to\infty} \frac{1}{l}\sum_{|s'|=l}H_l(s'),$$ where $|s'|=l$ indicates we are considering all the generated strings of length $l$. For a fixed string we can think on the normalized block entropy value where $l$ better captures the periodicity of $s$.

Entropy was originally conceived by Shannon as a measure of information transmitted over an stochastic communication channel with known alphabets and it establishes hard limits to maximum lossless compression rates. For instance, the Shannon coding (and Shannon-Fano) sorts the symbols of an alphabet according to their probabilities, assigning smaller binary self-delimited sequences to symbols that appear more frequently. Such methods form the base of many, if not most, commonly used compression algorithms.

Given its utility in data compression, entropy is often used as a measure of the information contained in a finite string $s=s_1s_2\dots{}x\dots{}s_k$. Let's consider the \textit{natural distribution}, the uniform distribution that makes the least number of assumptions but does consider every possibility equally likely and is thus uniform. Suggested by the set of symbols in $s$ and the string length the \textit{natural distribution} of $s$ is the distribution defined by $P(x)=\frac{n_x}{|s|}$, where $n_x$ is the number of times the object $x$ occurs in $s$ (at least one to be considered), and the respective entropy function $H_l$. If we consider blocks of size $n>>l\geq2$ and the string $s = 01010101\ldots01$, where $n$ is the length of the string, then $s$ can be compressed in a considerably smaller number of bits than a statistically random sequence of the same length and, correspondingly, has a lower $H_l$ value. However, entropy with the natural distribution suggested by the object, or any other computable distribution, is a computable function therefore is an imperfect approximation to algorithmic complexity.

\begin{figure}[ht!]
  \centering
 \includegraphics[width=1\textwidth]{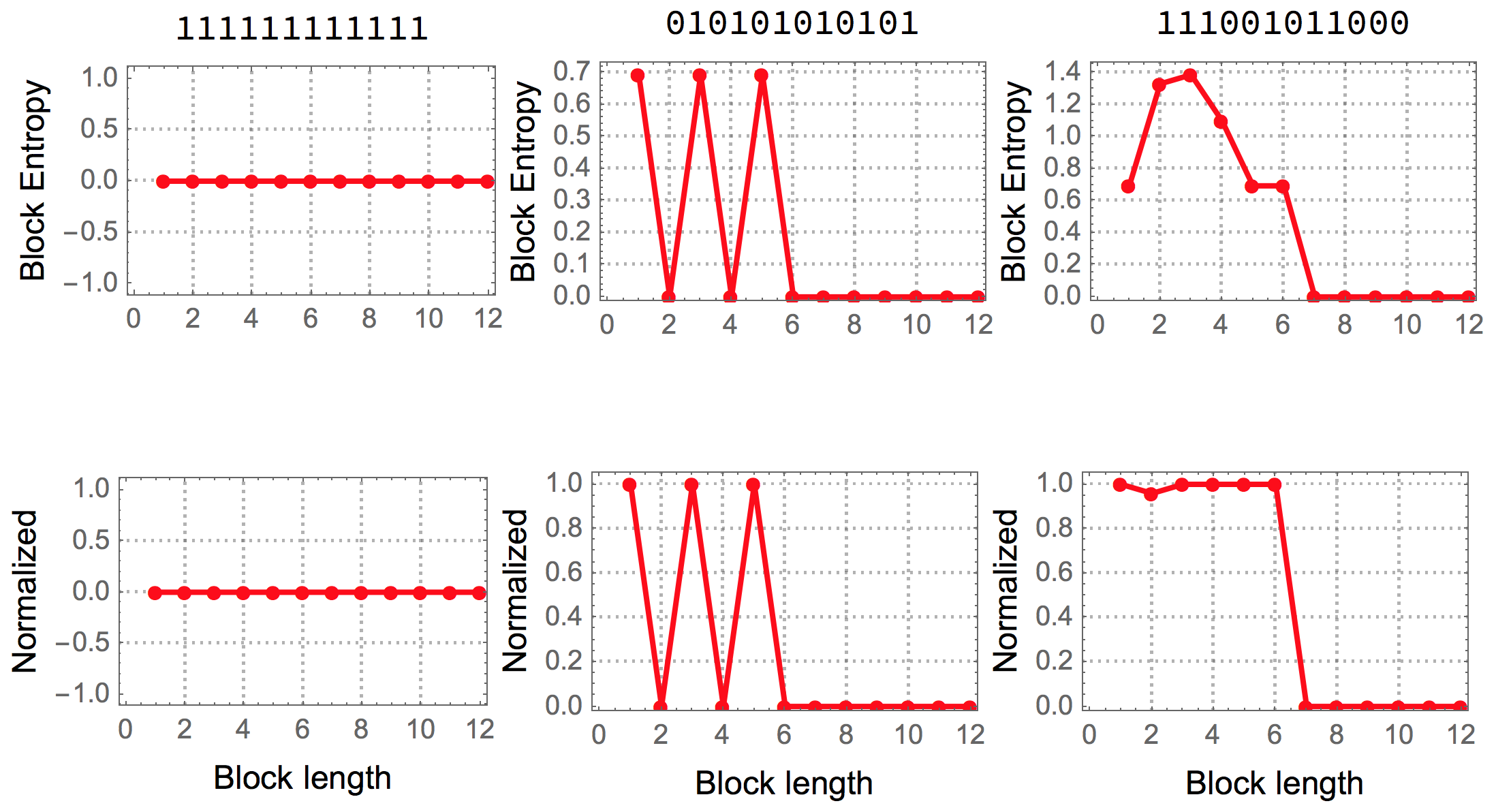}
   \caption{The best version of Shannon entropy can be rewritten as a function of variable block length where the minimum value best captures the (possible) periodicity of a string here illustrated with three strings of length 12, regular, periodic and random-looking. Because blocks larger than $n/2$ would result in only one block and therefore entropy equal to 0, the largest possible block is $n/2$. The normalized version (bottom) divides the entropy value for that block size by the largest possible number of blocks for that size and alphabet (here binary).}
  \label{blockentropy}
\end{figure}

The best possible version of a measure based upon entropy can be reached by partitioning an object into blocks of increasing size (up to half the length of the object) in order for Shannon entropy to capture any periodic statistical regularity. Fig.~\ref{blockentropy} illustrates the way in which such a measure operates on 3 different strings. 

\begin{figure}[ht!]
  \centering
 \includegraphics[width=11.5cm]{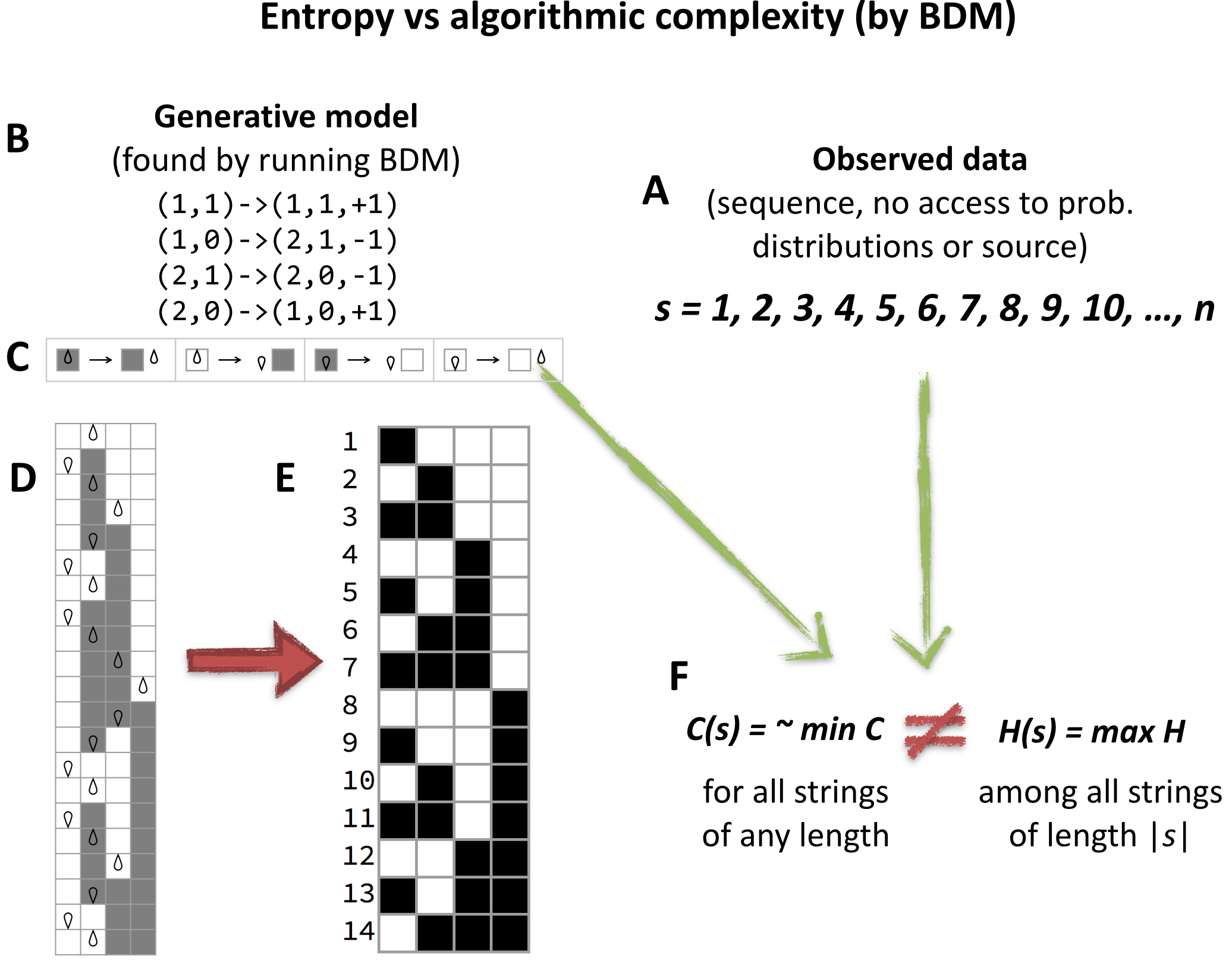}
   \caption{A: Observed data, a sequence of successive positive natural numbers. B: The transition table of the Turing machine found by running all possible small Turing machines. C: The same transition table in visual form. D: The space-time evolution of the Turing machine for starting from an empty tape. E: Space-time evolution of the Turing machine implementing a binary counter, taking as halting criterion the leftmost position of the original Turing machine head as depicted in C (states are arrows). E: This small computer program that our CTM and BDM methods find (c.f. next Section) mean that the sequence in A is not algorithmic random as the program represents a succinct generative causal model (and thus not random) for any arbitrary length that otherwise  would have been assigned a maximal randomness with Shannon Entropy  among all strings of the same length (in the face of no other knowledge about the source) despite its highly algorithmic non-random structured nature. Entropy alone---only equipped to spot statistical regularities when there is no access to probability distributions---cannot find this kind of generative models demonstrating the low randomness of an algorithmic sequence.}
  \label{binarycounter}
\end{figure}

However, no matter how sophisticated a version or variation of an Entropic measure will characterize certain algorithmic aspects of data that are not random but will appear to have maximal entropy if no knowledge about the source is known. In Fig.~\ref{binarycounter} depicted is how algorithmic probability/complexity can find such patterns and ultimately characterize any, including statistical ones, thereby offering a generalization and complementary improvement to the application of Entropy alone.

BDM builds upon block entropy's decomposition approach using algorithmic complexity methods to obtain and combine its building blocks. The result is a complexity measure that, as shown in section~\ref{boundproof}, approaches $K$ in the best case and behaves like entropy in the worst case (\ref{convergence-to-entropy}), outperforming $H_l$ in various scenarios. First we introduce the algorithm that conforms the building blocks of BDM, which are local estimations of algorithmic complexity. Specific examples of objects that not even Block entropy can characterize are found in Section~\ref{compcomp} showing how our methods are a significant improvement over any measure based upon entropy and traditional statistics. 

For example, the following two strings were assigned near maximal complexity but they were found to have low algorithmic complexity by CTM/BDM given that we were able to find not only a small Turing machine that reproduces them but also many Turing machines producing them upon halting thereby, by the Coding theorem, of low algorithmic complexity: 001010110101, 001101011010 (and their negations and reversions). These strings display nothing particularly special and they look in some sense typically random, yet this is what we were expecting, to find strings that would appear random but are actually not algorithmic random. This means that these strings would have been assigned higher randomness by Entropy and popular lossless compression algorithms but are assigned lower randomness when using our methods thus thus providing a real advantage over those other methods that can only exploit statistical regularities.

\section{The Coding Theorem Method (CTM)}

A computationally expensive procedure that is nevertheless closely related to algorithmic complexity involves approximating the algorithmic complexity of an object by running every possible program, from the shortest to the longest, and counting the number of times that a program produces every string object. The length of the computer program will be an upper bound of the algorithmic complexity of the object, following the Coding theorem (Eq.~\ref{codingeq}), and a (potentially) compressed version of the data itself (the shortest program found) for a given computer language or `reference' UTM. This guarantees discovery of the shortest program in the given language or reference UTM but entails an exhaustive search over the set of countable infinite computer programs that can be written in such a language. A program of length $n$ has asymptotic probability close to 1 of halting in time $2^n$~\cite{calude2}, making this procedure exponentially expensive, even assuming that all programs halt or that programs are assumed never to halt after a specified time, with those that do not being discarded. 

As shown in~\cite{delahayezenil} and~\cite{ctm}, an exhaustive search can be carried out for a small-enough number of computer programs (more specifically, Turing machines) for which the halting problem is known, thanks to the Busy Beaver problem~\cite{rado}. This problem consists in finding the Turing machine of fixed size (states and symbols) that runs longer than any other machine of the same size. Values are known for Turing machines with 2 symbols and up to 4 states that can be used to stop a resource-bounded exploration, that is, by discarding any machine taking more steps than the Busy Beaver values. For longer strings we also proceed with an informed runtime cut-off, below the theoretical $2^n$ optimal runtime that guarantees an asymptotic drop of non-halting machines~\cite{calude2} but above any value to capture most strings up to any degree of accuracy as performed in~\cite{zenilproofs}.

The so called \textit{Coding Theorem Method} (or simply CTM)~\cite{delahayezenil,ctm} is a bottom-up approach to algorithmic complexity and, unlike common implementations of lossless compression algorithms, the main motivation of CTM is to find algorithmic features in data rather than just statistical regularities that are beyond the range of application of Shannon entropy and popular lossless compression algorithms~\cite{emergence}. 

CTM is rooted in the relation~\cite{delahayezenil,ctm} provided by algorithmic probability between frequency of production of a string from a random program and its algorithmic complexity as described by Eq.~(\ref{codingeq}). Essentially it uses the fact that the more frequent a string is, the lower Kolmogorov complexity it has; and strings of lower frequency have higher Kolmogorov complexity. The advantage of using algorithmic probability to approximate $K$ by application of the Coding Theorem~\ref{codingeq} is that $m(s)$ produces reasonable approximations to $K$ based on an average frequency of production, which retrieves values even for small objects. 

Let $(t,k)$ denote the set of all Turing machines with $t$ states and $k$ symbols using the Busy Beaver formalism~\cite{rado}, and let $T$ be a Turing machine in $(t,k)$ with empty input. Then the empirical output distribution $D(t,k)$ for a sequence $s$ produced by some $T\in (t,k)$ gives an estimation of the \textit{algorithmic probability} of $s$, $D(t,k)(s)$ defined by:

\begin{definition}
\begin{equation}
\label{Deq} 
D(t,k)(s)=\frac{|\{T\in(t,k) : T \textit{ produces } s\}|}{|\{T \in(t,k) : T \textit{ halts }\}|}
\end{equation}
\end{definition}

For small values $t$ and $k$, $D(t,k)$ is computable for values of the Busy Beaver problem that are known. The Busy Beaver problem~\cite{rado} is the problem of finding the $t$-state, $k$-symbol Turing machine which writes a maximum number of non-blank symbols before halting, starting from an empty tape, or the Turing machine that performs a maximum number of steps before halting, having started on an initially blank tape. For $t=4$ and $k=2$, for example, the Busy Beaver machine has maximum runtime $S(t)=107$~\cite{brady}, from which one can deduce that if a Turing machine with 4 states and 2 symbols running on a blank tape hasn't halted after 107 steps, then it will never halt. This is how $D$ was initially calculated-- by using known Busy Beaver values. However, because of the undecidability of the Halting problem, the Busy Beaver problem is only computable for small $t,k$ values~\cite{rado}. Nevertheless, one can continue approximating $D$ for a greater number of states (and colours), proceeding by sampling, as described in~\cite{delahayezenil,ctm}, with an informed runtime based on both theoretical and numerical results. 

Notice that $0< D(t,k)(s) < 1$, $D(t,k)(s)$ and is thus said to be a semi-measure, just as $m(s)$ is. 

Now we can introduce a measure of complexity that is heavily reliant upon \textit{algorithmic probability} $m(s)$, as follows:

\begin{definition}
Let $(t,k)$ be the space of all $t$-state $k$-symbol Turing machines, $t,k > 1$ and $D(t,k)(s) =$ the function  assigned to every finite binary string $s$. Then:

\begin{equation}
CTM(s,t,k) = -\log_b D(t,k)(s)
\end{equation}

\noindent where $b$ is the number of symbols in the alphabet (traditionally 2 for binary objects, which we will take as understood hereafter).

That is, the more frequently a string is produced the lower its Kolmogorov complexity, with the converse also being true.
\end{definition}

Table~\ref{spacetable} shows the  rule spaces of Turing machines that were explored, from which empirical algorithmic probability distributions were sampled and estimated.

\begin{center}
	\begin{table}[h]
		\centering
		\resizebox{.8\textwidth}{!}{%
			\begin{tabular}{l|l|l|l}
				\textbf{(t,k)} & \textbf{Calculation} & \textbf{Number of Machines} & 
				\textbf{Time}\\
				\hline
				\hline
				(2,2) & $F$-- (6 steps) & $|R(2,2)| = 2000$   & 0.01 s \\
				\hline
				(3,2) & $F$-- (21) &  $|R(3,2)| = 2\,151\,296$  & 8 s \\
				(4,2) & $F$-- (107) &  $|R(4,2)| = 3\,673\,320\,192$ & 4 h\\
				\hline
				(4,2)$_{2D}$ & $F_{2D}$-- (1500) &  $|R(4,2)_{2D}| =
				315\,140\,100\,864$ & 252 d \\
				\hline
				(4,4) & $S$ (2000) & $334 \times 10^9$ & 62 d \\
				\hline
				(4,5) & $S$ (2000) & $214\times 10^9$  & 44 d \\
				\hline
				(4,6) & $S$ (2000) & $180\times 10^{9}$  & 41 d\\
				\hline
				(4,9) & $S$ (4000) & $200\times 10^{9}$ & 75 d  \\
				\hline
				(4,10) & $S$ (4000) & $201\times 10^{9}$  & 87 d \\
				\hline
				(5,2) & $F$-- (500) &  $|R(5,2)| = 9\,658\,153\,742\,336$  &
				450 d \\
				\hline
				(5,2)$_{2D}$ & $S_{2D}$ (2000) & $1291\times 10^9$  & 1970 d
			\end{tabular}%
			}
		\caption{\label{spacetable}Calculated empirical distributions from rulespace $(t,k)$. Letter codes: $F$ full space, $S$ sample, $R(t,k)$ reduced enumeration. Time is given in seconds (s), hours (h) and days (d).}
	\end{table}
\end{center}
\begin{definition}
We will designate as \textit{base string}, \textit{base matrix}, or \textit{base tensor} the objects of size $l$ for which CTM values were calculated such that the full set of $k^l$ objects have CTM evaluations. In other words, the base object is the maximum granularity of application of CTM.
\end{definition}

Table~\ref{spacetable} provides figures relating to the number of base objects calculated.\\

Validations of CTM undertaken before show the correspondence between CTM values and the exact number of instructions used by Turing machines when running to calculate CTM~\cite{computability}(Fig 1 and Table 1) to produce each string, i.e. direct $K$ complexity values for this model of computation (as opposed to CTM using algorithmic probability and the Coding theorem) under the chosen model of computation~\cite{rado}. The correspondence in values found between the directly calculated $K$ and CTM by way of frequency production was near perfect. 

Sections 7.1.2 and 7.2, and Figs. 10, 11, 12 and 15 in~\cite{kolmo2d} support the agreements in correlation using different rule spaces of Turing machines and different computing models altogether (cellular automata). Section 7.1.1 of the same paper provides a first comparison to lossless compression. The sections `Agreement in probability' and `Agreement in rank' provide further material comparing rule space (5,2) to the rule space (4,3) previously calculated in~\cite{delahayezenil}. The section `Robustness' in~\cite{ctm} provides evidence relating to the behaviour of the \textit{invariance theorem constant} for a standard model of Turing machines~\cite{rado}.

\section{The Block Decomposition Method (BDM)}
\label{ctm}

Because finding the program that reproduces a large object is computationally very expensive and ultimately uncomputable, one can aim at finding short programs that reproduce small fragments of the original object, parts that together compose the larger object. And this is what the BDM does. 

BDM is divided into two parts. On the one hand, approximations to $K$ are performed by CTM which values can then be used and applied in $O(1)$ by exchanging time for memory in the population of a precomputed look-up table for small strings, which would diminish its precision as a function of object size (string length) unless a new iteration of CTM is precomputed again. On the other hand, BDM decomposes the original data into fragments for which CTM provides an estimation and then puts the values together based upon classical information theory.

BDM is thus a hybrid complexity measure that combines Shannon Entropy in the long range but provides local estimations of algorithmic complexity. It is meant to improve the properties of Shannon Entropy that in practice are reduced to finding statistical regularities and to extend the power of CTM. It consists in decomposing objects into smaller pieces for which algorithmic complexity approximations have been numerically estimated using CTM, then reconstructing an approximation of the Kolmogorov complexity for the larger object by adding the complexity of the individual components of the object, according to the rules of information theory. For example, if $s$ is an object and $10s$ is a repetition of $s$ ten times smaller, upper bounds can be achieved by approximating $K(s)+\log_2(10)$ rather than $K(10s)$, because we know that repetitions have a very low Kolmogorov complexity, given that one can describe repetitions with a short algorithm. 

Here we introduce and study the properties of this \emph{Block Decomposition Method} based on a method advanced in~\cite{delahayezenil,ctm} that takes advantage of the powerful relationship established by algorithmic probability between the frequency of a string produced by a random program running on a (\emph{prefix-free}) UTM and the string's Kolmogorov complexity. The chief advantage of this method is that it deals with small objects with ease, and it has shown stability in the face of changes of formalism, producing reasonable Kolmogorov complexity approximations. 
BDM must be combined with CTM if it is to scale up properly and behave optimally for upper bounded estimations of $K$. BDM $+$ CTM is universal in the sense that it is guaranteed to converge to $K$ due to the invariance theorem, and as we will prove later, if CTM no longer runs, then BDM alone approximates the Shannon entropy of a finite object.

 Like compression algorithms, BDM is subject to a trade-off. Compression algorithms deal with the trade-off of compression power and compression/decompression speed. 

\subsection{\textit{l}-overlapping String Block Decomposition}

Let us fix values for $t$ and $k$ and let $D(t,k)$ be the frequency distribution constructed from running all the Turing machines with $n$ states and $k$ symbols. Following Eq.~(\ref{Deq}), we have it that $-\log D$ is an approximation of $K$ (denoted by $CTM$). We define the BDM of a string or finite sequence $s$ as follows,\\

\begin{definition}
\begin{equation}
\label{ecaeq}
BDM(s, l, m) = \sum_i CTM(s^i, m, k) + \log(n_i)
\end{equation}

\noindent where $n_i$ is the multiplicity of $s^i$ and $s^i$ the subsequence $i$ after decomposition of $s$ into subsequences $s^i$, each of length $l$, with a possible remainder sequence $y < |l|$ if $|s|$ is not a multiple of the decomposition length $l$.
\end{definition}

The parameter $m$ goes from 1 to the maximum string length produced by CTM, where $m=l$ means no overlapping inducing a partition of the decomposition of $s$, $m$ is thus an overlapping parameter when $m<l$ for which we will investigate its impact on BDM (in general, the smaller $m$ a greater overestimation of BDM).

The parameter $m$ is needed because of the remainder. If  $|s|$ is not a multiple of the decomposition length $l$ then the option is to either ignore the remainder in the calculation of BDM or define a sliding window with overlapping $m-l$.

The choice of $t$ and $k$ for CTM in BDM depend only on the available resources for running CTM, which involves running the entire $(t,k)$ space of Turing machines with $t$ symbols and $k$ states.

BDM approximates $K$ in the following way: if $p_i$ is the minimum program that generates each base string $s^i$, then $CTM(s^i) \approx |p_i|$ and we can define an unique program $q$ that runs each $p_i$, obtaining all the building blocks. How many times each block is present in $s$ can be given in $O(\log n_i)$ bits. Therefore, BDM is the sum of the information needed to describe the decomposition of $s$ in base strings. How close is this sum to $K$ is explored in section \ref{boundproof}.

The definition of BDM is interesting because one can plug in other algorithmic distributions, even computable ones, approximating some measure of algorithmic complexity even if it is not the one defined by Kolmogorov-Chaitin such as, for example, the one defined by Calude et al.~\cite{calude} based upon Finite-state automata. BDM thus allows the combination of measures of classical information theory and algorithmic complexity.

For example, for binary strings we can use $t=2$ and $k=2$ to produce the empirical output distribution $(2,2)$ of all machines with 2 symbols and 2 states by which all strings of size $l=12$ are produced, except two (one string and its complement). But we assign them values $\max{\{CTM(y,2,2)+r : |y|=12\}}$ where $e$ is different from zero because the missing strings were not generated in $(2,2)$ and therefore have a greater algorithmic random complexity than any other string produced in $(2,2)$ of the same length. Then, for $l = 12$ and $m = 1$, $BDM(s,l,m)$ decomposes $s=010101010101010101$ of length $|s|=18$ into the following subsequences:\\

\begin{center}
010101010101\\
101010101010\\
010101010101\\
101010101010\\
010101010101\\
101010101010\\
010101010101\\
\end{center}

\noindent with 010101010101 having multiplicity 4 and 101010101010 multiplicity 3.

We then get the CTM values for these sequences:

\begin{center}
$CTM(010101010101,2,2) = 26.99073$\\

$CTM(101010101010,2,2) = 26.99073$
\end{center}

To calculate BDM, we then take the sum of the CTM values plus the sum of the $\log_b$ of the multiplicities, with $b=2$ because the string alphabet is 2, the same as the number of symbols in the set of Turing machines producing the strings. Thus:

$$\log_2(3) + \log_2(4) + 26.99 + 26.99 = 57.566$$

\subsection{2- and \textit{w}-Dimensional Complexity}

To ask after the likelihood of an array, we can consider a 2-dimensional Turing machine. The Block Decomposition Method can then be extended to objects beyond the unidimensionality of strings, e.g. arrays representing bitmaps such as images, or graphs (by way of their adjacency matrices). We would first need CTM values for 2- and $w$-dimensional objects that we call base objects (e.g. base strings or base matrices).

\begin{figure}[ht!]
  \centering
  \includegraphics[width=9.5cm]{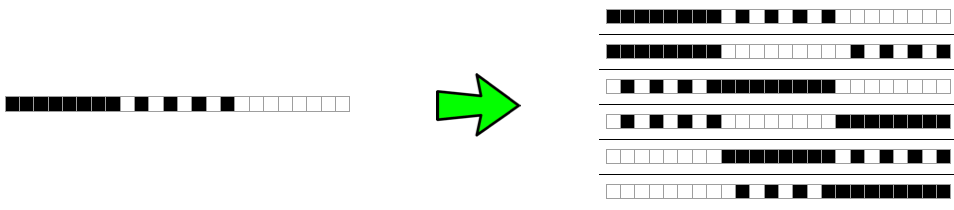}\\

\bigskip
\bigskip

  \includegraphics[width=10.5cm]{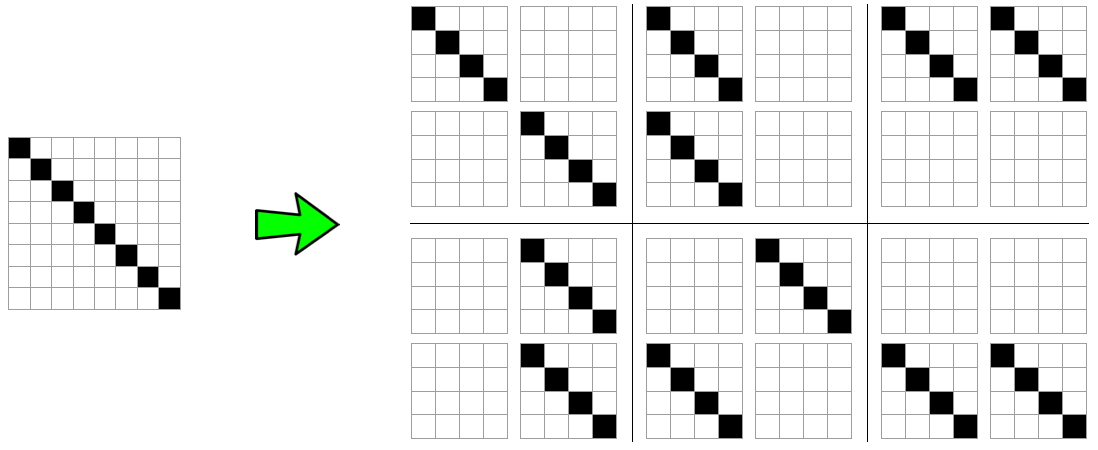}
   \caption{Non-overlapping BDM calculations are invariant to block permutations (reshuffling base strings and matrices), even when these permutations may have different complexities due to the reorganization of the blocks that can produce statistical or algorithmic patterns. For example, starting from a string of size 24 (top) or an array of size $8\times8$ (bottom), with decomposition length $l=8$ for strings and decomposition $l=4\times 4$ block size for the array, all 6 permutations for the string and all 6 permutations for the array have the same BDM value regardless of the shuffling procedure.}
  \label{permutations}
\end{figure}

A popular example of a 2-dimensional tape Turing machine is Langton's ant~\cite{langton}. Another way to see this approach is to take the BDM as a way of deploying all possible 2-dimensional deterministic Turing machines of a small size in order to reconstruct the adjacency matrix of a graph from scratch (or smaller pieces that fully reconstruct it). Then, as with the \textit{Coding theorem method} (above), the algorithmic complexity of the adjacency matrix of the graph can be estimated via the frequency with which it is produced from running random programs on the (prefix-free) 2-dimensional Turing machine. More specifically,

\begin{equation}\label{BDM}
BDM(X,\{x_i\}) = \displaystyle \sum_{(r_i,n_i)\in{}Adj(X)_{\{x_i\}}}
CTM(r_i) + \text{log} (n_i),
\end{equation}
\noindent{}where the set $Adj(X)_{\{x_i\}}$ is composed of the pairs 
$(r,n)$, $r$ is an element of the decomposition of $X$
(as specified by a \textit{partition} $\{x_i\}$, where $x_i$ is a submatrix of $X$) in different sub-arrays of size up to $d_1
\times \ldots \times d_w$ (where $w$ is the dimension of the object) that we call \textit{base matrix} (because $CTM$ values were obtained for them) and $n$ is the multiplicity of each component. $CTM(r)$ is a computable approximation from below to the algorithmic information complexity of $r$, $K(r)$, as obtained by applying the coding theorem method to $w$-dimensional Turing machines. In other words, $\{r_i\}$ is the set of \textit{base objects}. 

Because string block decomposition is a special case of matrix block decomposition, and square matrix block decomposition is a special case of $w$-block decomposition for objects of $w$ dimensions, let us describe the way in which BDM deals with boundaries on square matrices, for which we can assume CTM values are known, and that we call base strings or base matrices. 

Fig.~\ref{permutations} shows that the number of permutations is a function of the complexity of the original object, with the number of permutations growing in proportion to the original object's entropy--because the number of different resulting blocks determines the number of different $n$ objects to distribute among the size of the original object (e.g. 3 among 3 in Fig.~\ref{permutations} (top) or only 2 different $4 \times 4$ blocks in Fig.~\ref{permutations} (bottom)). This means that the non-overlapping version of BDM is not invariant vis-\`a-vis the variation of the entropy of the object, on account of which it has a different impact on the error introduced in the estimation of the algorithmic complexity of the object. Thus, non-overlapping objects of low complexity will have little impact, but with random objects non-overlapping increases inaccuracy. Overlapping decomposition solves this particular permutation issue by decreasing the number of possible permutations, in order to avoid trivial assignment of the same BDM values. However, overlapping has the undesired effect of systematically overestimating values of algorithmic complexity by counting almost every object of size $n$, $n-1$ times, hence overestimating at a rate of about $n(n-1)$ for high complexity objects of which the block multiplicity will be low, and by $n\log(n)$ for low complexity objects. 

Applications to graph theory~\cite{zenilgraph}, image classification~\cite{gauvrit1} and human behavioural complexity have been produced in the last few years~\cite{gauvrit2,kempe}.

\subsection{BDM Upper and Lower Absolute Bounds}
\label{boundproof}

In what follows we show the hybrid nature of the measure. We do this by setting lower and upper bounds to BDM in terms of the algorithmic complexity $K(X)$, the partition size and the approximation error of $CTM$, such that these bounds are tighter in direct relation to smaller partitions and more accurate approximations of $K$. These bounds are independent of the partition strategy defined by $\{x_i\}$.

\begin{prop}\label{t1}
    
    Let $BDM$ be the function defined in Eq.~\ref{BDM} and let $X$ be an array of dimension $w$. Then $K(X) \leq BDM(X,\{x_i\}) + O(\log ^2 |A|) + \epsilon$ and $BDM(X,\{x_i\}) \leq |Adj(X)_{\{x_i\}}|K(X) + O(|Adj(X)_{\{x_i\}}|\log |Adj(X)_{\{x_i\}}|) - \epsilon$, where $A$ is a set composed of all possible ways of accommodating 
    the elements of $Adj(X)_{\{x_i\}}$ in an array of dimension $w$, and $\epsilon$ is the sum of errors for the approximation $CTM$ over all the sub-arrays used.

    \begin{proof}
        
        Let $Adj(X)_{\{x_i\}} = \{(r_1,n_1),...,(r_k,n_k)\}$ and
        $\{p_j\}$, $\{t_j\}$ be the sequences of programs for the reference prefix-free UTM $U$ such that, for
        each $(r_j,n_j) \in Adj(X)_{\{x_i\}}$, we have $U(p_j) = r_j$,
        $U(t_j)=n_j$, $K(r_j) = |p_j|$ and $|t_j| \leq 2 \log (n_j) + c$.
        Let $\epsilon_j$ be a positive constant such that 
        $CTM(r_j) + \epsilon_j = K(X)$; this is the error for each sub-array. Let $\epsilon$ be the sum of all the errors. 
        
        For the first inequality we can construct a program $q_w$, whose 
        description only depends on $w$, such that, given a description of 
        the set $Adj(X)_{\{x_i\}}$ and an index $l$, it enumerates all the 
        ways of accommodating the elements in the set and returns the array 
        corresponding to the position given by $l$. 
        
        Note that $|l|$, $|Adj(X)_{\{x_i\}}|$ and all $n_j$'s are of the 
        order of $\text{log} |A|$. Therefore 
        $U(q_wq_1p_1t_1...p_jt_jl) = X$ and
        \begin{align*}
        K(X) &\leq  |q_wp_1t_1...p_jt_jl| \\
        &\leq  |q_w| + \displaystyle\sum_{1}^k (|q_j| + |p_j|) + |l|\\
        &\leq BDM(X,\{x_i\}) + \epsilon + |q_w| \\ 
        &+(\log |A| + c)|Adj(X)_{\{x_i\}}| + O(\log |A|) \\
        &\leq  BDM(X,\{x_i\}) + O(\text{log}^2 |A|) + \epsilon,
        \end{align*}
        \noindent{}which gives us the inequality. 
        
        Now, let $q_X$ be the smallest program that generates $X$. For the second inequality we can describe a program $q_{\{x_i\}}$ which, given a description of $X$ and the index $j$, constructs the set $Adj(X)_{\{x_i\}}$ and returns $r_j$, i.e. $U(q_{\{x_i\}}q_Xj)=r_j$. Note that each $|j|$ is of the order of
        $\log |Adj(X)_{\{x_i\}}|$. Therefore, for each $j$ we have 
        $K(r_j) + \epsilon_j = |p_j| \leq
        |q_{\{x_i\}}| +  |q_X| + O(\log |Adj(X)_{\{x_i\}}|)$
        and $K(r_j) + \epsilon _j + \log (n_i) \leq
        |q_{\{x_i\}}| + |q_X| + O(\log |Adj(X)_{\{x_i\}}|) + \log (n_i)$. 
        Finally, by adding all the terms over the $j$'s we find
        the second inequality:
        \begin{align*}
        &BDM(X,\{x_i\})+ \epsilon \leq |Adj(X)_{\{x_i\}}|(|q_X|+|q_{\{x_i\}}| \\&+\log(n_j)+O(\log |Adj(X)_{\{x_i\}}|))\leq|Adj(X)_{\{x_i\}}|K(X)\\
        &+O(|Adj(X)_{\{x_i\}}|\log|Adj(X)_{\{x_i\}}|).
        \end{align*}
    \end{proof}
\end{prop}
\begin{cor}\label{t2} 
    If the partition defined by $\{x_i\}$ is small, that is, if 
    $|Adj(X)_{\{x_i\}}|$ is close to 1, then $BDM(X,\{x_i\}) \approx
    K(X)$.
    
    \begin{proof}
        Given the inequalities presented in proposition \ref{t1}, we 
        have it that 
        
        \begin{equation*}
        K(X) - O(\log ^2 |A|) - \epsilon \leq
        BDM(M,\{x_i\})
        \end{equation*}
        and
        \begin{equation*}
        \begin{split}
        &BDM(M,\{x_i\}) \leq \\
        &|Adj(X)_{\{x_i\}}|K(X)
        +O(|Adj(X)_{\{x_i\}}|\log |Adj(X)_{\{x_i\}}|) + \epsilon
        \end{split}
        \end{equation*}
        which at the limit leads to $K(X) - \epsilon \leq BDM(X) \leq K(X) - \epsilon$ and 
        $BDM(X) = K(X) - \epsilon$. From~\cite{ctm}, we can say that the error rate $\epsilon$ is small, and that by the invariance theorem it will converge towards a constant value.
    \end{proof}
\end{cor}

\section{Dealing with Object Boundaries}

Because partitioning an object---a string, array or tensor---leads to boundary leftovers not multiple of the partition length, the only two options to take into consideration such boundaries in the estimation of the algorithmic complexity of the entire object is to either estimate the complexity of the leftovers or to define a sliding window allowing overlapping in order to include the leftovers in some of the block partitions. The former implies mixing object dimensions that may be incompatible (e.g. CTM complexity based on 1-dimensional TMs versus CTM based on higher dimensional TMs). Here we explore these strategies to deal with the object boundaries. Here we introduce a strategy for partition minimization and \textit{base object} size maximization that we will illustrate for 2-dimensionality. The strategies are intended to overcome under- or over-fitting complexity estimations that are due to conventions, not just  technical limitations (due to, e.g., uncomputability and intractability). 

Notice that some of the explorations in this section may give the wrong impression to use and introduce ad-hoc methods to deal with the object boundaries. However, this is not the case. What we will do in this section is to explore all possible ways we could conceive in which we can estimate $K$ according to BDM taking into considerations the boundaries that may require special treatment when they are not of multiple size to the partition length from the decomposition of the data after BDM. Moreover, we show that in all cases, the results are robust because the errors found are convergent and can thus be corrected, so any possible apparently ad-hoc condition has little to no implications in the calculation of BDM in the limit and only very limited at the beginning.

\subsection{Recursive BDM}

In Section~\ref{boundproof}, we have shown that using smaller partitions for $BDM$ yields more accurate approximations to the algorithmic complexity $K$. However, the computational costs for calculating $CTM$ are high. We have compiled an exhaustive database for square matrices of size up to $4 \times 4$. Therefore it is in our best interest to find a method to minimize the partition of a given matrix into squares of size up to $d\times{}d=l$ for a given $l$. 

The strategy consists in taking the biggest base matrix multiple of $d\times{}d$ on one corner and dividing it into adjacent square submatrices of the given size. Then we group the remaining cells into 2 submatrices and apply the same procedure, but now for $(d-1)\times{}(d-1)$. We continue dividing into submatrices of 
size $1\times{}1$. 

Let $X$ be a matrix of size $m \times n$ with $m,n \geq d$. Let's denote by $\text{quad}= \{UL, LL, DR, LR\}$ the set of quadrants on a matrix and by $\text{quad}^d$ the set of vectors of quadrants of dimension $l$. We define a function $part(X,d,q_i)$, where $\langle q_1, \ldots ,q_d\rangle \in
\text{quad}^d$, as follows:
\begin{equation*}
\begin{split}
part(X,l,q_i) =&max(X,d,q_i)\\
&\cup part(resL(X,d,q_i), d-1, q_{i+1})\\
&\cup part(resR(X,d,q_i), d-1, q_{i+1})\\ 
&\cup part(resLR(X,d,q_i), d-1, q_{i+1}),
\end{split}
\end{equation*}
\noindent{}where $max(X,d,q_i)$ is the largest set of adjacent submatrices of size $d \times d$ that can be clustered in the corner corresponding to the quadrant $q_i$, $resR(X,d-1,q_i)$ is the submatrix composed of all the adjacent rightmost cells that could not fit on $max(X,d,q_i)$ and are not part of the leftmost cells, $resL(X,d-1,q_i)$ is an analogue for the leftmost cells and $resLR(X,d-1,q_i)$ is the submatrix composed of the cells belonging to the rightmost and leftmost cells. We call the last three submatrices \emph{residual matrices}.

By symmetry, the number of matrices generated by the function is invariant with respect to any vector of quadrants $\langle q_1, \ldots, q_d\rangle$. However, the final BDM value can (and will) vary according to the partition chosen. Nevertheless, with this strategy we can evaluate all the possible BDM values for a given partition size and choose the partition that yields the minimum value, the maximum value, or compute the average for all possible partitions. 

The partition strategy described can easily be generalized and applied to strings (1 dimension) and tensors (objects of $n$-dimensions).

\subsection{Periodic Boundary Conditions}

One way to avoid having remaining matrices (from strings to tensors) of different sizes is to embed a matrix in a topological torus (see Fig.~\ref{torus} bottom) such that no more object borders are found. Then let $X$ be a square matrix of arbitrary size $m$. We screen the matrix $X$ for all possible combinations to minimize the number of partitions maximizing block size. We then take the combination of smallest BDM for fixed \textit{base matrix} size $d$ and we repeat for $d-1$ until we have added all the components of the decomposed $X$. This procedure, will, however, overestimate the complexity values of all objects (in unequal fashion along the complexity spectra) but will remain bounded, as we will show in Section~\ref{error}.

Without loss of generality the strategy can be applied to strings (1 dimension) and tensors (any larger number of dimensions, e.g. greater than 2), the former embedded in a cylinder while tensors can be embedded in $n$-dimensional tori (see Fig.~\ref{torus}).

\section{BDM versus Shannon Entropy}
\label{compcomp}

Let us address the task of quantifying how many strings with maximum entropy rate are actually algorithmically compressible, i.e., have low algorithmic complexity. That is, how many strings are actually algorithmically (as opposed to simply statistically) compressible but are not compressed by lossless compression algorithms, which are statistical (entropy rate) estimators~\cite{emergence}. We know that most strings have both maximal entropy (most strings look equally statistically disordered, a fact that constitutes the foundation of thermodynamics) and maximal algorithmic complexity (according to a pigeonhole argument, most binary strings cannot be matched to shorter computer programs as these are also binary strings). But the gap between those with maximal entropy and low algorithmic randomness diverges and is infinite at the limit (for an unbounded string sequence). That is, there is an infinite number of sequences that have maximal entropy but low algorithmic complexity.

\begin{figure}[ht!]
\centering
\footnotesize{\hspace{-3.5cm}\textbf{A}\hspace{6cm}\textbf{B}}\\

\scalebox{.23}{\includegraphics{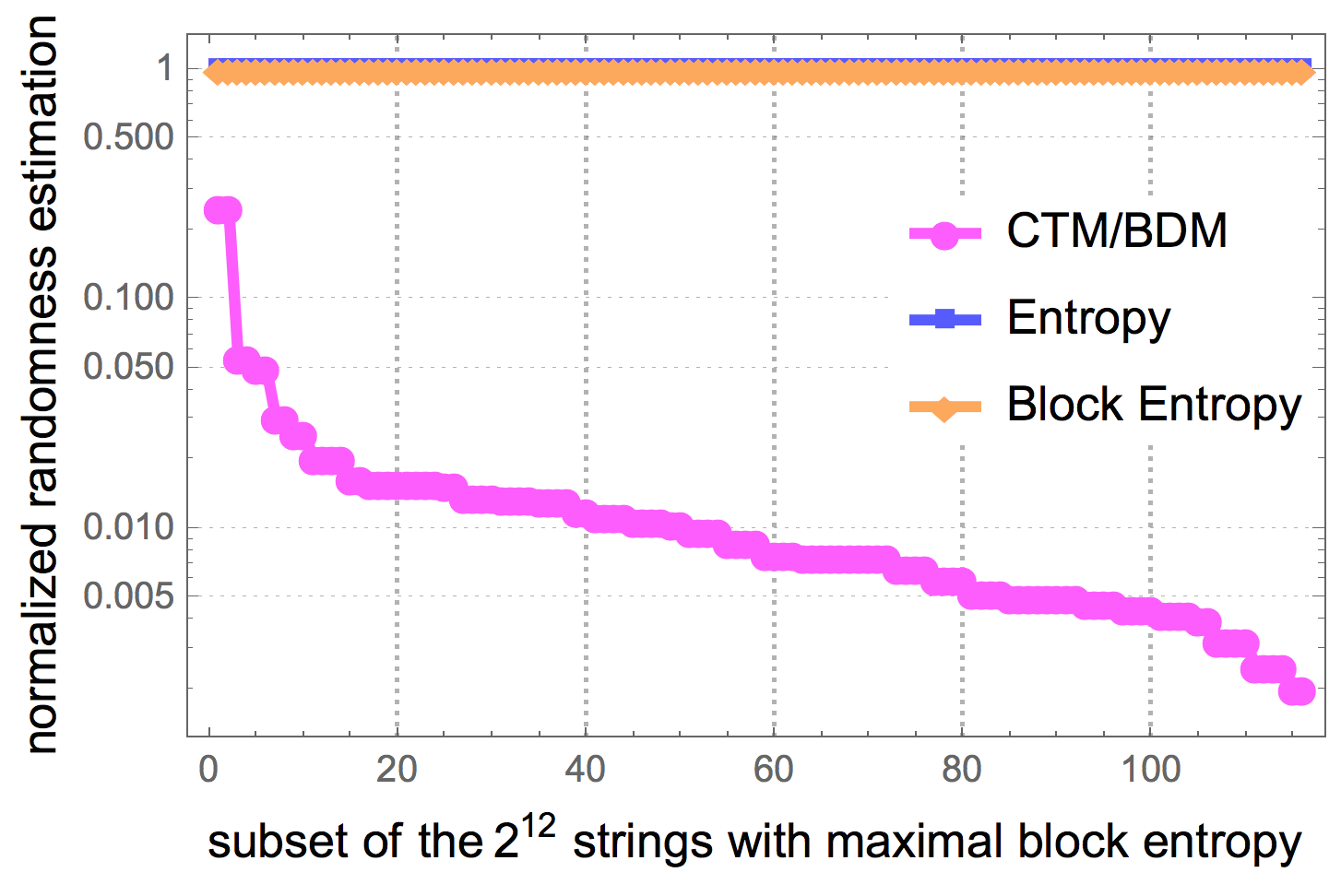}}\hspace{1cm}\scalebox{.14}{\includegraphics{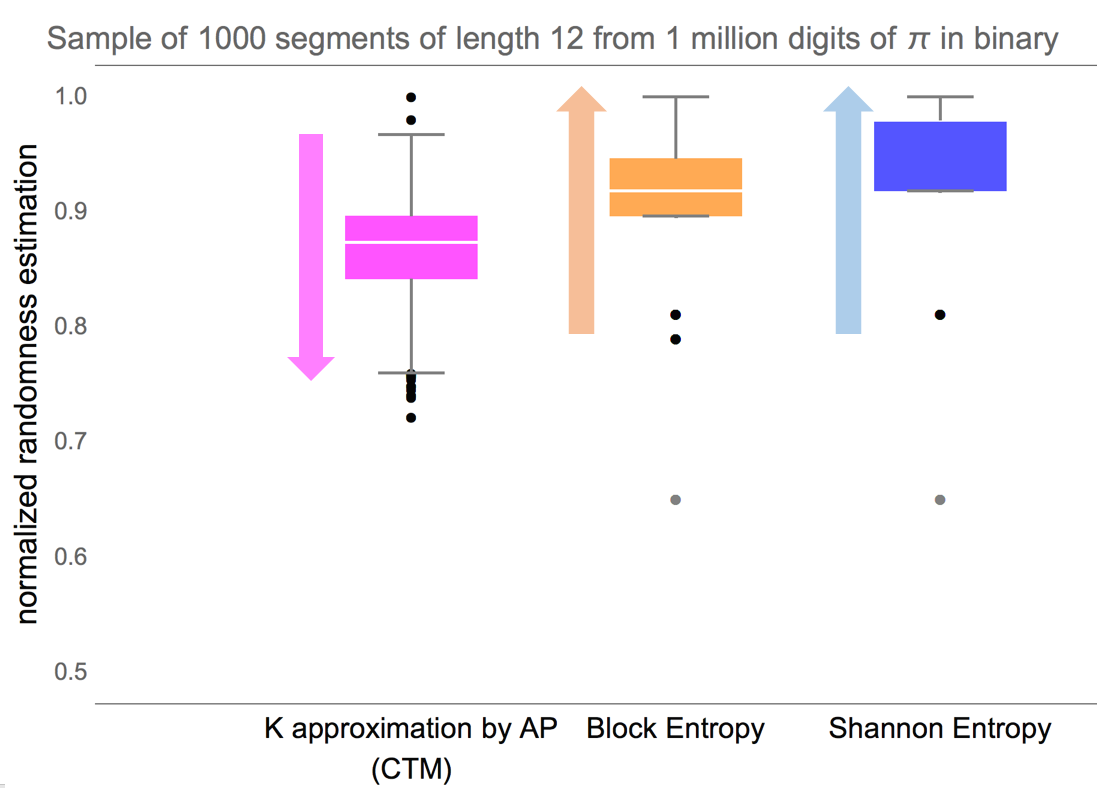}}\\

\footnotesize{\hspace{-3.5cm}\textbf{C}\hspace{6cm}\textbf{D}}\\

\scalebox{.215}{\includegraphics{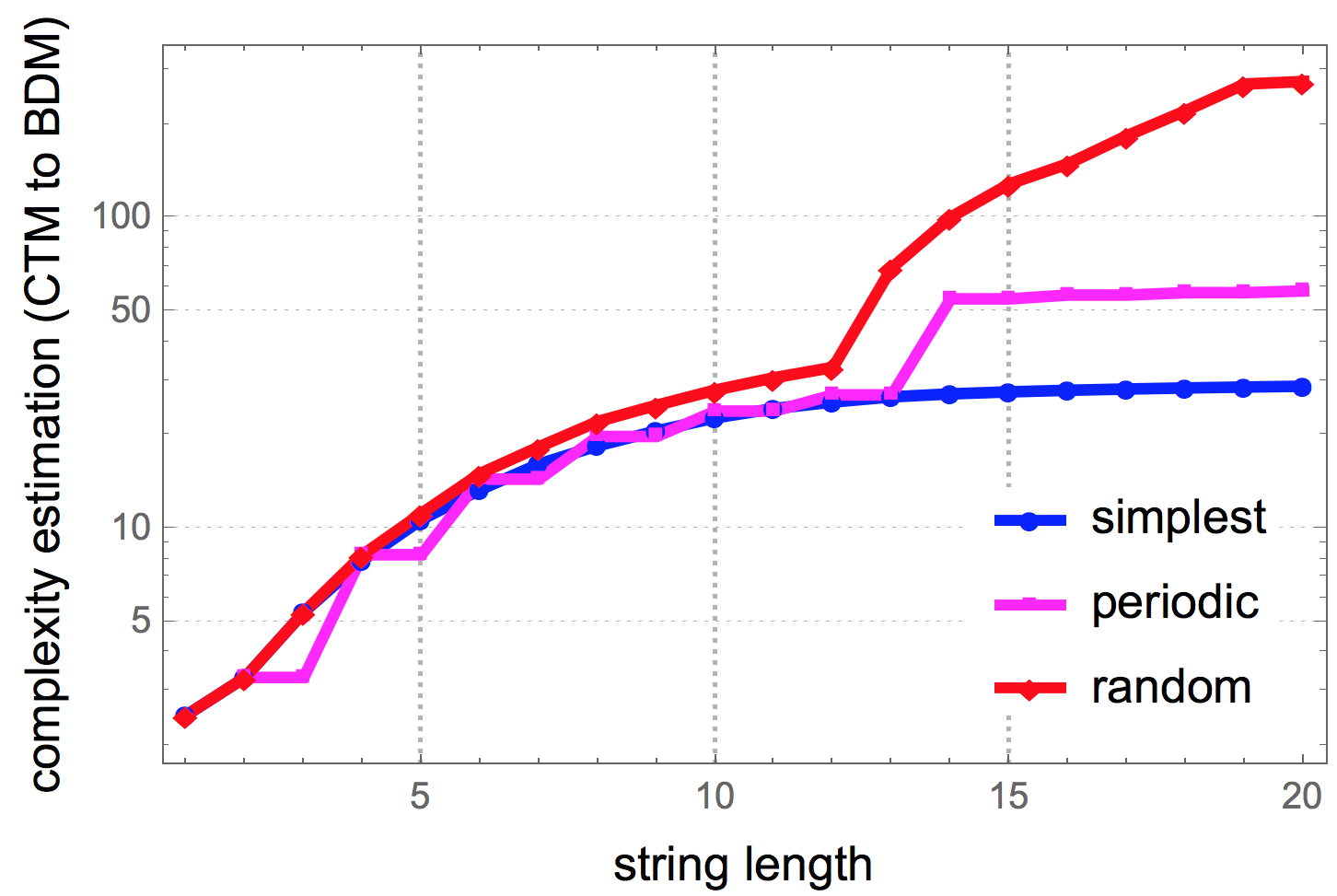}}\hspace{1cm}\scalebox{.215}{\includegraphics{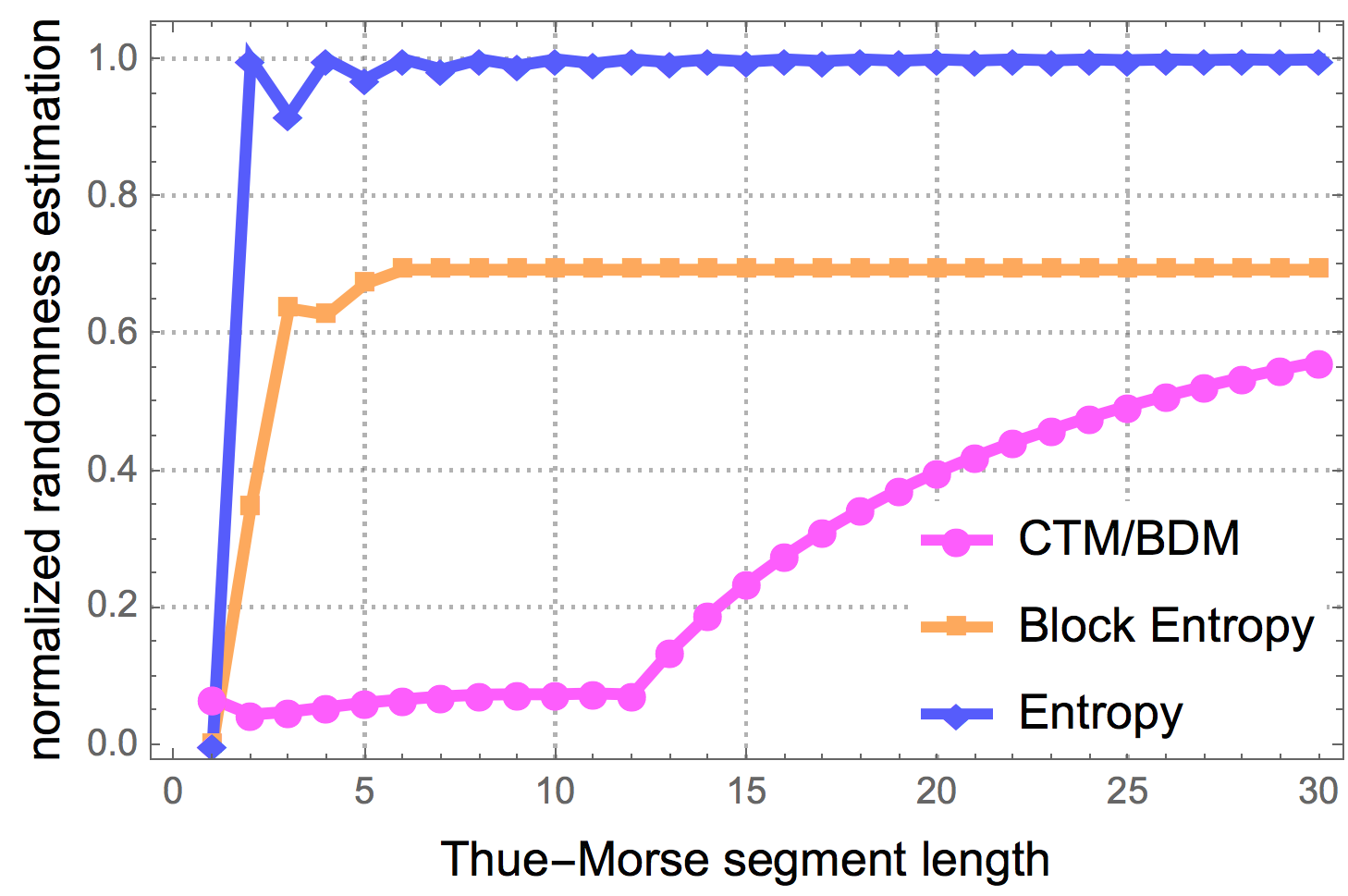}}\\

\footnotesize{\hspace{-3.5cm}\textbf{E}\hspace{6cm}\textbf{F}}\\

\scalebox{.21}{\includegraphics{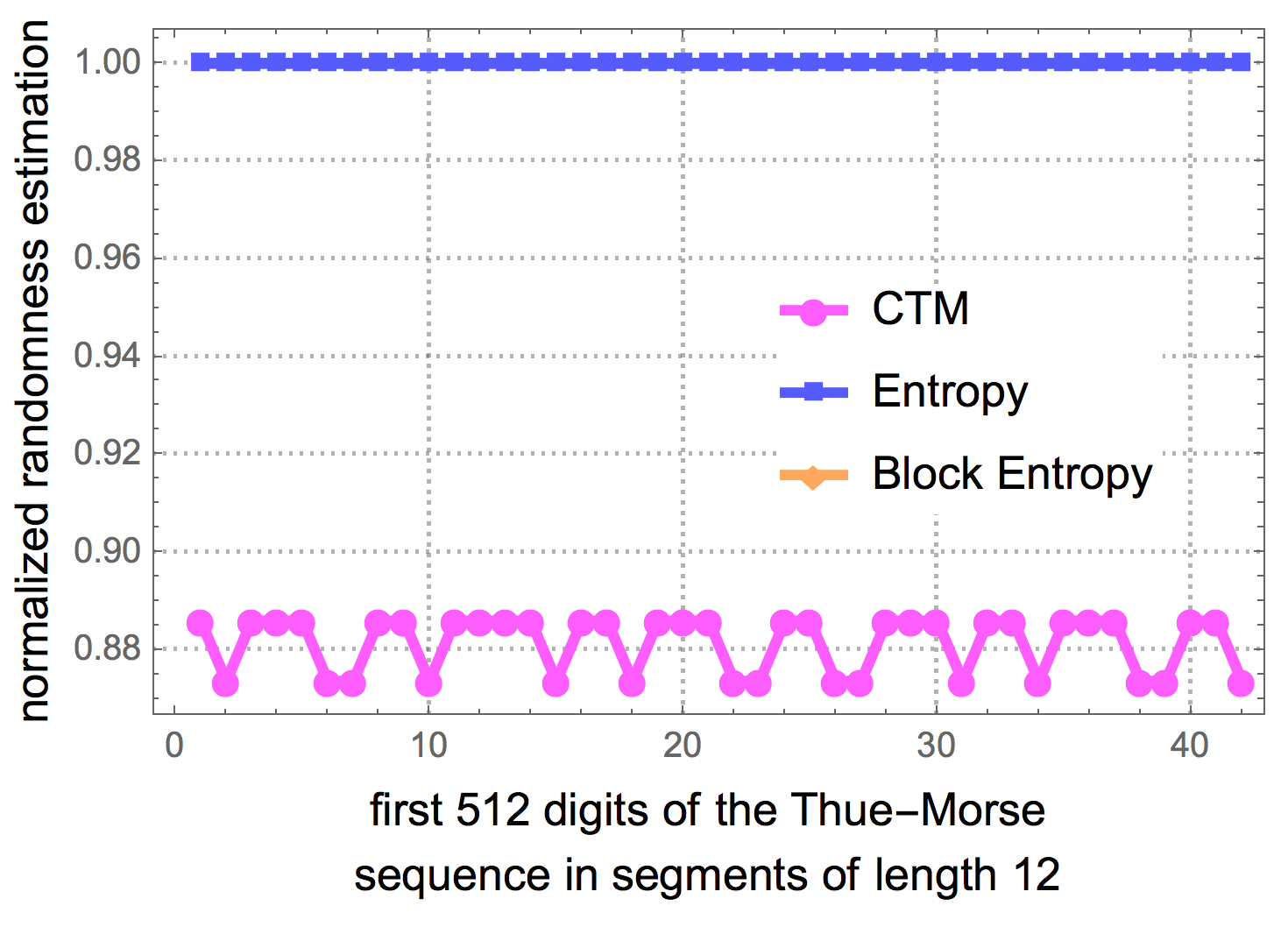}}\hspace{1cm}\scalebox{.21}{\includegraphics{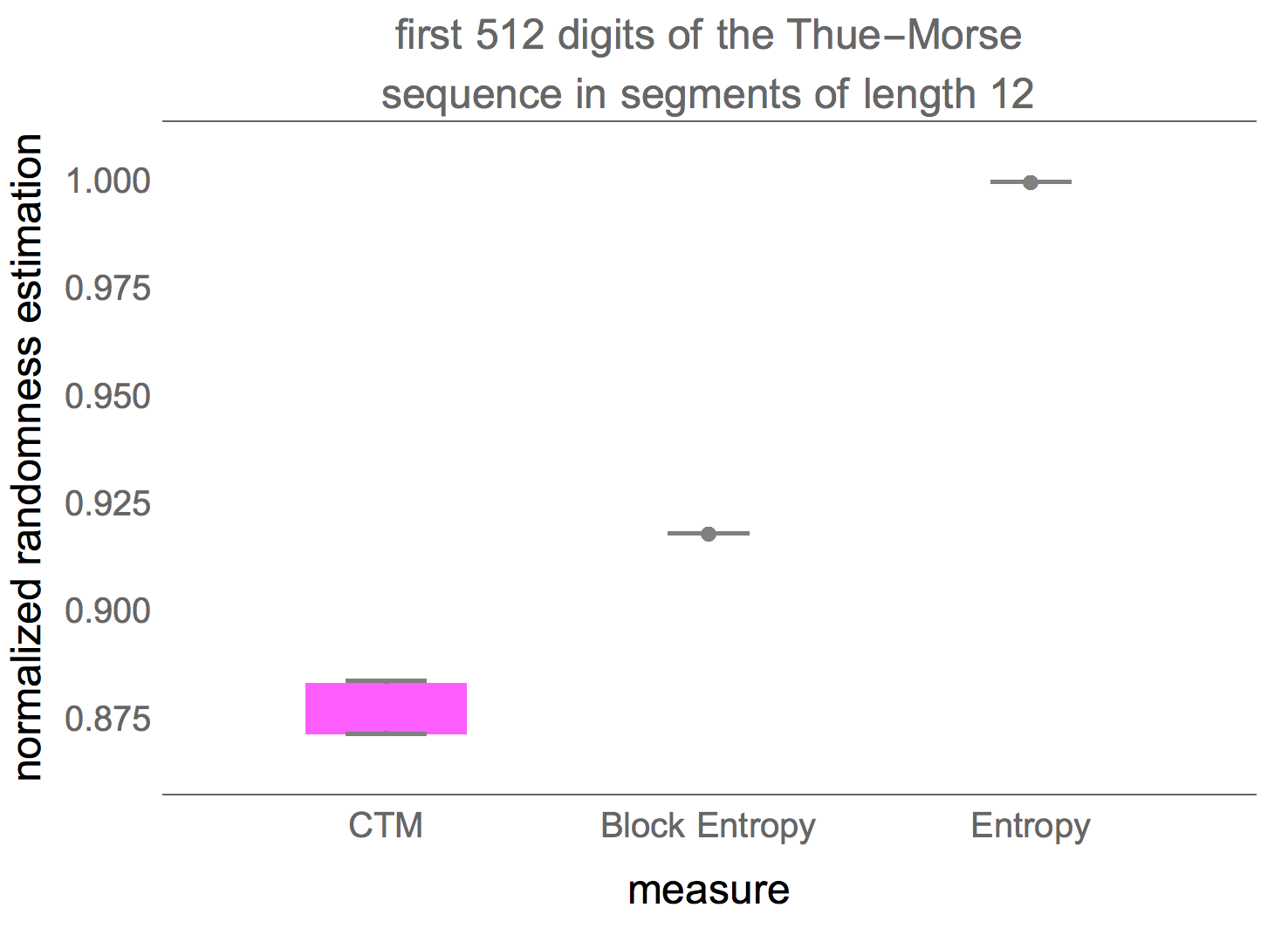}}\\

\caption{\label{piandthue}Telling $\pi$ and the Thue-Morse sequence apart from truly (algorithmic) random sequences. CTM assigns significantly lower randomness (B, D, E and F) to known low algorithmic complexity objects. (B) If absolute Borel normal (as strongly suspected and statistically demonstrated to any confidence degree), $\pi$'s entropy and block entropy asymptotically approximate 1 while, by the invariance theorem of algorithmic complexity, CTM asymptotically approximates 0. Smooth transitions between CTM and BDM are also shown (C and D) as a function of string complexity. Other smooth transition functions of BDM are explored and introduced in Subsection~\ref{mibdm}.}
\end{figure}

\begin{figure}[htbp!]
  \centering
\includegraphics[width=0.4\textwidth]{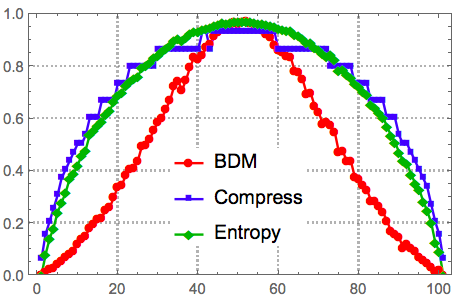}\hspace{1cm} \includegraphics[width=0.4\textwidth]{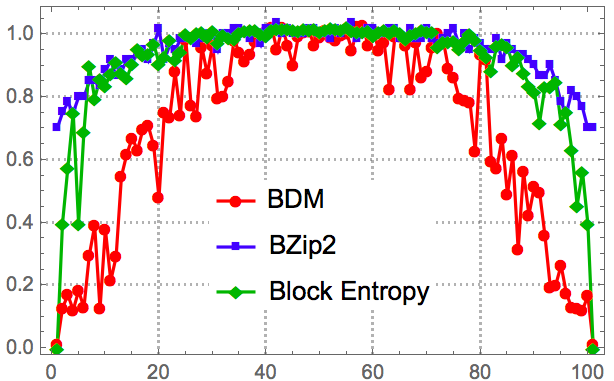}\\

\bigskip
\bigskip

\includegraphics[width=0.4\textwidth]{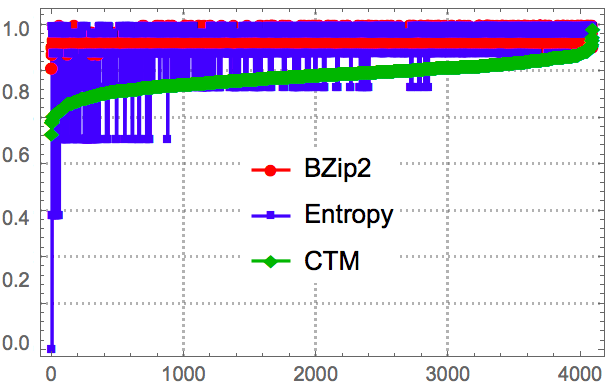}\hspace{1cm}\includegraphics[width=0.4\textwidth]{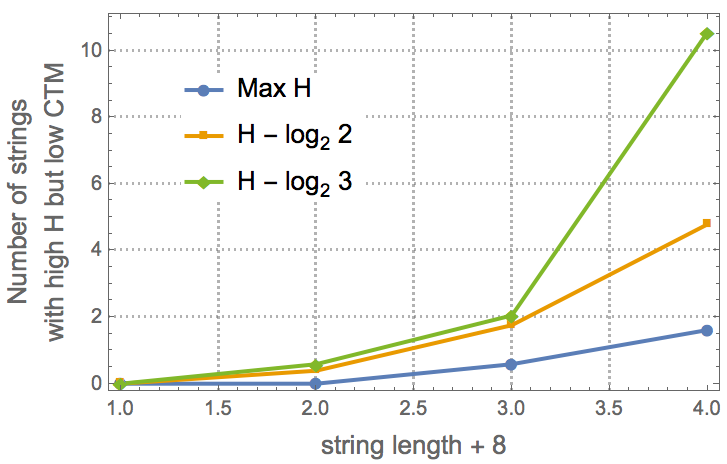}
  \caption{\label{minK}Strings that are assigned lower randomness than that estimated by entropy. Top left: Comparison between values of entropy, compression (\texttt{Compress[]}) and BDM over a sample of 100 strings of length 10\,000 generated from a binary random variable following a Bernoulli distribution and normalized by maximal complexity values. Entropy follows a Bernoulli distribution and, unlike compression that follows entropy, BDM values produce clear convex-shaped gaps on each side assigning lower complexity to some strings compared to both entropy and compression. Top right: The results confirmed using a popular lossless compression algorithm BZip2 (and also confirmed, even if not reported, with LZMA) on 100 random strings of 100 bits each (BZip2 is slower than Compress but achieves greater compression). Bottom left: The $CTM_{low}(s) - H_{high}(s)$ gap between near-maximal entropy and low algorithmic complexity grows and is consistent along different string lengths, here from 8 to 12 bits. This gap is the one exploited by BDM and carried out over longer strings, which gives it the algorithmic edge against entropy and compression. Bottom right: When strings are sorted by CTM, one notices that BZip2 collapses most strings to minimal compressibility. Over all $2^{12} = 4\,096$ possible binary strings of length 12, entropy only produces 6 different entropy values, but CTM is much more fine-grained, and this is extended to the longer strings by BDM, which succeeds in identifying strings of lower algorithmic complexity that have near-maximal entropy and therefore no statistical regularities. Examples of such strings are in Section~\ref{compcomp}.}
\end{figure}

The promise of BDM is that, unlike compression, it does identify some cases of strings with maximum entropy that actually have low algorithmic complexity. Figs.~\ref{piandthue} and ~\ref{minK} shows that indeed BDM assigns lower complexity to more strings than entropy, as expected. Unlike entropy, and implementations of lossless compression algorithms, BDM recognizes some strings that have no statistical regularities but have algorithmic content that makes them algorithmically compressible.

Examples of strings with lower randomness than that assigned by entropy and Block entropy are 101010010101 (and its complement) or the Thue-Morse sequence 011010011001$\ldots$ (or its complement) obtained by starting with 0 and successively appending the Boolean complement~\cite{morse}, the first with low CTM $= 29$ and the Thue-Morse with CTM $= 33.13$ (the max CTM value in the subset is $37.4$ for the last string in this table). The Morse sequence is uniformly recurrent without being periodic, not even eventually periodic, so will remain with high entropy and Block entropy.

CTM and BDM as functions of the object's size (and therefore the size of the Turing machine rule space that has to be explored) have the following time complexity:

\begin{myitemize}
\item CTM is uncomputable but for decidable cases runs in $O(exp)$ time.
\item Non-overlapping string BDM and LD runs in $O(1)$ linear time and $O(n^d)$ polynomial time for $d$-dimensional objects. 
\item Overlapping BDM runs in $O(ns)$ time with $m$ the overlapping offset.
\item Full overlapping with $m=1$ runs in $O(2^n)$ polynomial time as a function of the number of overlapping elements $n$.
\item Smooth BDM runs in $O(1)$ linear time.
\item Mutual Information BDM runs in $O(exp)$ time for strings and $d$ exponential for dimension $d$.
\end{myitemize}

So how does this translate into real profiling power of recursive strings and sequences that are of low algorithmic complexity but appear random to classical information theory? Fig.~\ref{piandthue} provides real examples showing how BDM can outperform the best versions of Shannon entropy.

Fig.~\ref{piandthue} shows the randomness estimation of 2 known low algorithmic complexity objects and CTM to BDM transitions of the mathematical constant $\pi$ and the Thue-Morse sequence both of which numerical estimations by CTM assign lower randomness than the suggested by both entropy and its best version Block entropy. It is expected to find that CTM does much better at characterizing the low algorithmic randomness of a sequence like the Thue-Morse sequence (beyond the fact that it is not Borel normal~\cite{morse}) given that every part of the sequence is algorithmically obtained from another part of the sequence (by logical negation or the substitution system 0$\rightarrow$01, 1$\rightarrow$10 starting from 0) while the digits of $\pi$ have been shown to be independent (at least in powers of 2) from each other~\cite{bbp} and only algorithmic in the way they are produced from any of the many known generating formulas. 

One of the most recent found formula producing any digits of any segment of the mathematical constant $\pi$ (in base $2^k$, is given by a very short symbolic summation~\cite{bbp}:
$\sum_{n=1}^{\infty} (4/(8 n + 1) - 2/(8 n + 4) - 1/(8 n + 5) - 1/(8 n + 6))/k^n$
To produce such a string of 80 ASCII characters (less than 1K bits) has a probability of $1/256^{83} \times f$ or $1.30642\times 10^{-200} \times f$ to be produced by chance typing `random formulae', where $f$ is a multiplying factor quantifying the number other formulas of fixed (small) size that can also produce $\pi$ and for which there are many known since the works of Vieta, Leibniz, Wallis, Euler, Ramanujan, and others. In theory, classical probability is exponentially divergent from the much higher algorithmic probability $1/2^n$ that is the classical probability to produce an initial segment of $\pi$ (in binary) of length $n$. A good source of formulae producing digits of $\pi$ can be found at e.g. the Online Encyclopedia of Integer Sequences (OEIS) (\url{https://oeis.org/A000796}) listing more than 50 references and at Wolfram MathWorld listing around a hundred (\url{http://mathworld.wolfram.com/PiFormulas.html}).

Unlike classical probability, algorithmic probability quantifies the production of the object (Fig.~\ref{piandthue}, and also Fig. \ref{supfig5} in the Appendix) by indirect algorithmic/recursive means rather than by direct production (the typical analogy is writing on a computer equipped with a computer language compiler program versus writing on a typewriter).

To the authors knowledge, no other numerical method (see Fig.~\ref{piandthue}, and also Fig. \ref{supfig5} in the Appendix) is known to suggest low algorithmic randomness characterizations of statistical random-looking constants such as $\pi$ and the Thue-Morse sequence from an observer perspective with no access to prior information, probability distributions or knowledge about the nature of the source (i.e. a priori deterministic). 

There are 2 sides of BDM. On the one hand, it is based on a non-computable function but once approximations are computed we build a lookup table of values which makes BDM computable. Lossless compression is also computable but it is taken as able to make estimations of an uncomputable function like $K$ because they can provide upper bounds and estimate $K$ from above just as we do with CTM (and thus BDM) by exhibiting a short Turing machine capable of reproducing the data/string.

\begin{center}
\footnotesize{
	\begin{table}[h]
	\caption{\label{range}Summary of ranges of application and scalability of CTM and all versions of BDM. $d$ stands for the dimension of the object.}
		\begin{center}
			\tabcolsep=0.11cm
			{\footnotesize
			\resizebox{.6\textwidth}{!}{%
			\begin{tabular}{c|c|c|c}
				& \textbf{short strings}  & \textbf{long strings} & \\
				& $< 100$ bits & $>100$ bits & \textbf{scalability}\\
				\hline
				\hline
				Lossless  &  &  & \\
				compression  & $\times$ & \ding{51} & $O(n)$\\
				\hline
				\textit{Coding Theorem}  &  & &\\
				 \textit{Method (CTM)}  & \ding{51} & $\times$ & $O(exp)$ to $O(\infty)$\\
				\hline
				\textit{Non-overlapping}  &  & & \\
				\textit{BDM}  & \ding{51} &\ding{51} & $O(n)$\\
				\hline
				\textit{Full-overlapping} &  & & \\
				 \textit{Recursive BDM} &  \ding{51} & \ding{51} & $O(n^{d-1})$\\
				\hline
				\textit{Full-overlapping}  &  & & \\
				 \textit{Smooth BDM} & \ding{51} & \ding{51} & $O(n^{d-1})$\\
				\hline
				\textit{Smooth add} &  & & \\
				\textit{col BDM}  & \ding{51} &\ding{51} & $O(n)$
			\end{tabular}%
			}
			}
		\end{center}
		\end{table}
		}
\end{center}

Table~\ref{range}, summarizes the range of application, with CTM and BDM preeminent in that they can more efficiently deal with short, medium and long sequences and other objects such as graphs, images, networks and higher dimensionality objects.

\section{Error Estimations}
\label{error}

One can estimate the error in different calculations of BDM, regardless of the error estimations of CTM (quantified in~\cite{delahayezenil,ctm}), in order to calculate their departure and deviation both from granular entropy and algorithmic complexity, for which we know lower and upper bounds. For example, a maximum upper bound for binary strings is the length of the strings themselves. This is because no string can have an algorithmic complexity greater than its length, simply because the shortest computer program (in bits) to produce the string may be the string itself. 

\begin{figure}[ht!]
\centering
\scalebox{.3}{\includegraphics{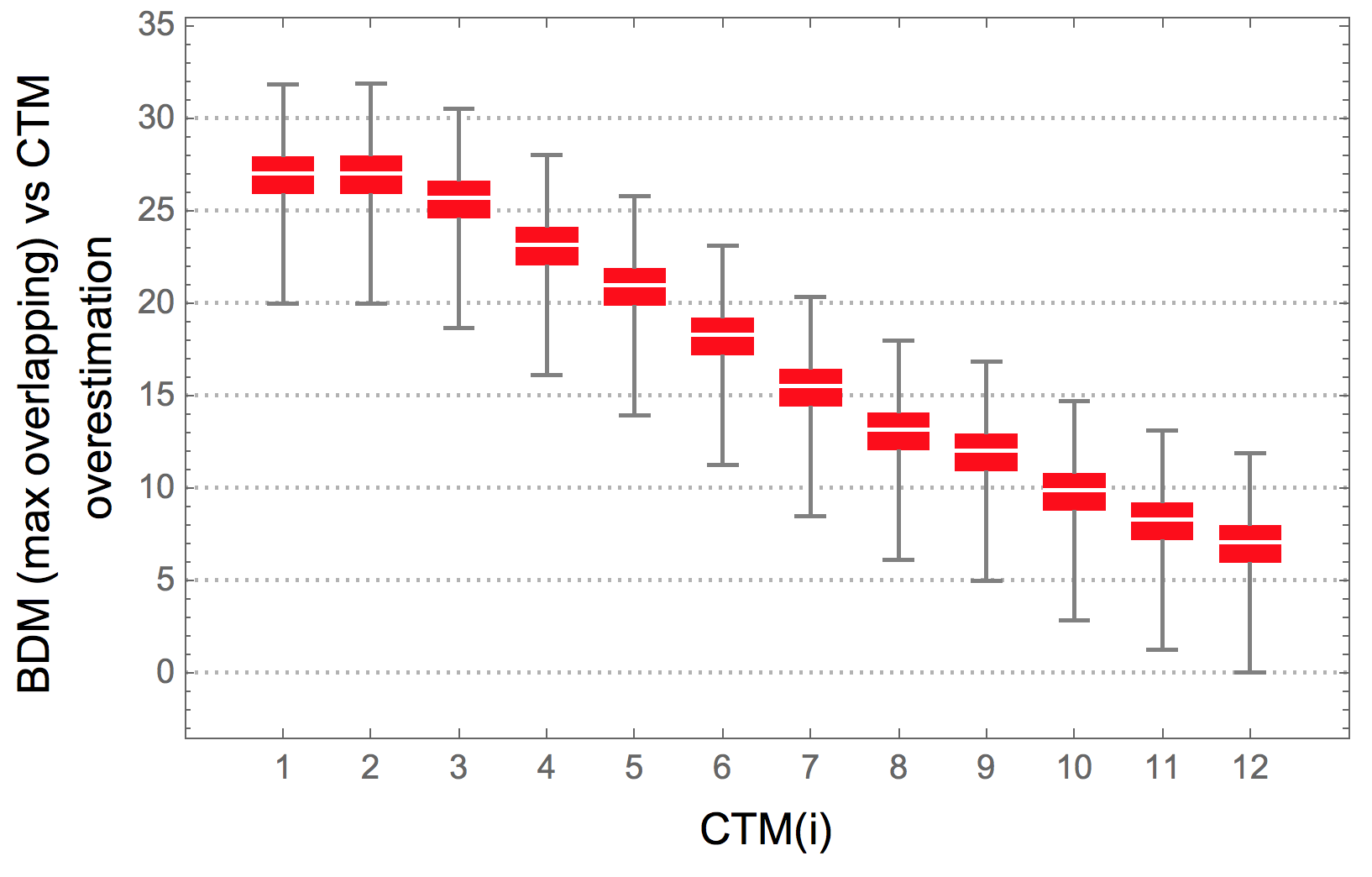}}\\

\caption{\label{ctmbdmerror}Box plot showing the error introduced by BDM quantified by CTM. The sigmoid appearance comes from the fact that we actually have exact values for CTM up to bitstrings of length 12 and so $BDM(12) = CTM(i)$ for $i = 12$ but the slope of the curve gives an indication of the errors when assuming that BDM has only access to $CTM(i)$ for $i < 12$ versus actual $CTM(12)$. This means that the error grows linearly as a function of CTM and of the string length, and the accuracy degrades smoothly and slowly towards entropy if CTM is not updated.}
\end{figure}

In the calculation of BDM, when an object's size is not a multiple of the base object of size $d$, boundaries of size $<d$ will be produced, and there are various ways of dealing with them to arrive at a more accurate calculation of an object that is not a multiple of the base. First we will estimate the error introduced by ignoring the boundaries or dealing with them in various ways, and then we will offer alternatives to take into consideration in the final estimation of their complexity.

\begin{figure}[ht!]
\centering
\includegraphics[width=0.95\textwidth]{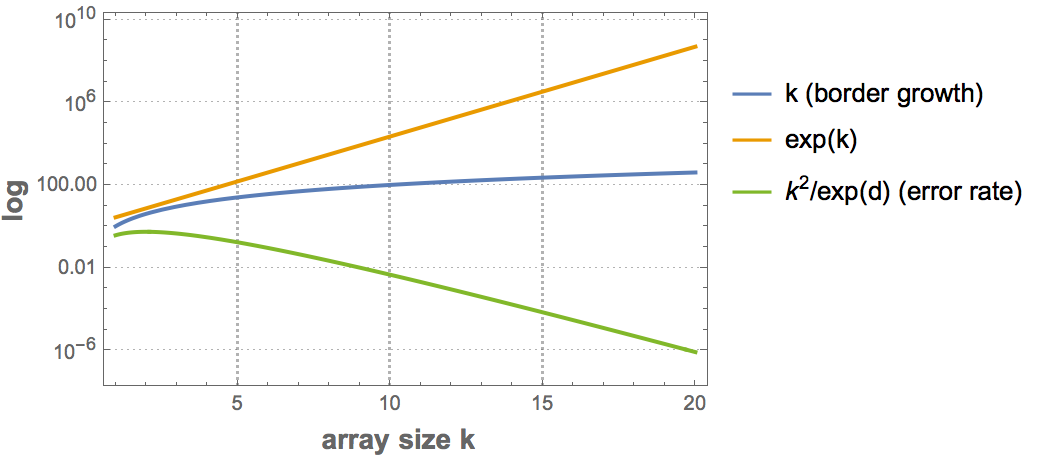}\\

\caption{\label{errorplot}Error rate for 2-dimensional arrays. With no loss of generalization, the error rate for $n$-dimensional tensors $\lim_{ d\rightarrow \infty} \frac{k^n}{n^k}=0$ is convergent and thus negligible, even for the discontinuities disregarded in this plot which are introduced by some BDM versions, such as non-overlapping blocks and discontinuities related to trimming the boundary condition.}
\end{figure}

If a matrix $X$ of size $k\times{}j$ is not a multiple of the base matrix of size $d\times{}d$, it can be divided into a set of decomposed blocks of size $d\times d$, and $R$, $L$, $T$ and $B$ residual matrices on the right, left, top and bottom boundaries of $M$, all of smaller size than $d$.

\begin{figure}[ht!]
\centering
\scalebox{.25}{\includegraphics{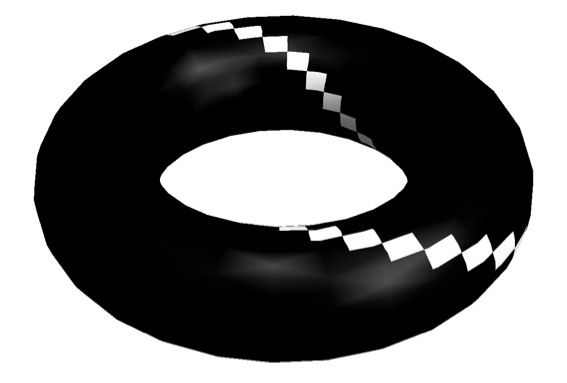}} \scalebox{.25}{\includegraphics{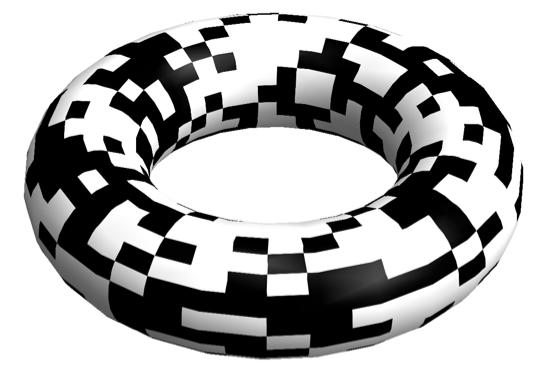}} \scalebox{.25}{\includegraphics{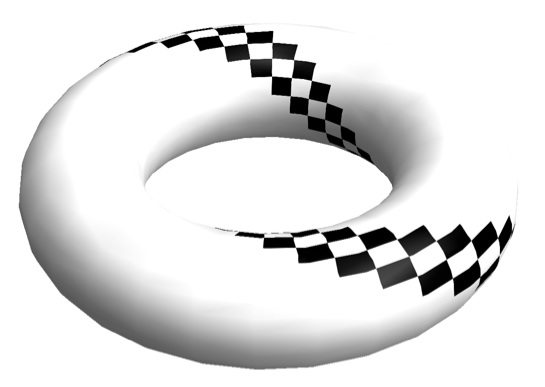}}
\caption{\label{torus}One way to deal with the decomposition of $n$-dimensional tensors is to embed them in an $n$-dimensional torus ($n=2$ in the case of the one depicted here), making the borders cyclic or periodic by joining the borders of the object. Depicted here are three examples of graph canonical adjacency matrices embedded in a 2-dimensional torus that preserves the object complexity on the surface, a complete graph, a cycle graph and an Erd\"os-R\'enyi graph with edge density 0.5, all of size 20 nodes and free of self-loops. Avoiding borders has the desired effect of producing no residual matrices after the block decomposition with overlapping.}
\end{figure}

Then boundaries $R$, $L$, $T$ and $B$ can be dealt with in the following way:

\begin{myitemize}
\item Trimming boundary condition: $R$, $L$, $T$ and $B$ are ignored, then $BDM(X)=BDM(X, R, L, T, B)$, with the undesired effect of general underestimation for objects not multiples of $d$. The  error introduced (see Fig.~\ref{error}) is bounded between 0 (for matrices divisible by $d$) and $k^2/exp(k)$, where $k$ is the size of $X$. The error is thus convergent ($exp(k)$ grows much faster than $k^2$) and can therefore be corrected, and is negligible as a function of array size as shown in Fig.~\ref{error}.
\item Cyclic boundary condition (Fig.~\ref{torus} bottom): The matrix is mapped onto the surface of a torus such that there are no more boundaries and the application of the overlapping BDM version takes into consideration every part of the object. This will produce an over-estimation of the complexity of the object but will for the most part respect the ranking order of estimations if the same overlapping values are used with maximum overestimation $d-1\times \max\{CTM(b)|b \in X\}$, where $K(b)$ is the maximum CTM value among all base matrices $b$ in $X$ after thedecomposition of $X$.
\item Full overlapping recursive decomposition: $X$ is decomposed into $(d-1)^2$ base matrices of size $d \times d$ by traversing $X$ with a sliding square block of size $d$. This will produce a polynomial overestimation in the size of the object of up to $(d-1)^2$, but if consistently applied it will for the most part preserve ranking.
\item Adding low complexity rows and columns (we call this `add col'): If a matrix of interest is not multiple the base matrices, we add rows and columns until completion to the next multiple of the base matrix, then we correct the final result by substracting the borders that were artificially added.
\end{myitemize}

The BDM error rate (see~\ref{torus} top) is the discrepancy of the sum of the complexity of the missed borders, which is an additive value of, at most, polynomial growth. The error is not even for a different complexity. For a tensor of $d$ dimensions, with all 1s as entries, the error is bounded by $\log(k^d)$ for objects with low algorithmic randomness and by $\frac{k^d}{d^k}$ for objects with high algorithmic randomness. 

Ultimately there is no optimal strategy for making the error disappear, but in some cases the error can be better estimated and corrected~\ref{error} and all cases are convergent, hence asymptotically negligible, and in all cases complexity ranking is preserved and under- and over-estimations bounded.

\subsection{BDM Worse-case Convergence towards Shannon Entropy}
\label{convergence-to-entropy}

Let $\{x_i\}$ be a partition of $X$ defined as in the previous sections for a fixed $d$. Then the Shannon entropy of $X$ for the partition 
$\{x_i\}$ is given by:
\begin{equation}
H_{\{x_i\}}(X) =-\displaystyle\sum_{(r_j,n_j)\in
Adj(X)_{\{x_i\}}}\frac{n_j}{|\{x_i\}|}
\log (\frac{n_j}{|\{x_i\}|}),
\end{equation}
where $P(r_j)=\frac{n_j}{|\{x_i\}|}$ and the array $r_j$ is taken as a symbol itself.
The following proposition establishes the asymptotic relationship between $H_{\{x_i\}}$ and $BDM$.

\begin{prop}\label{t3}
Let $M$ be a 2-dimensional matrix and $\{x_i\}$ a partition 
strategy with elements of maximum size $d \times d$. Then:
\begin{equation*}
|BDM_{\{x_i\}}(X)-H_{\{x_i\}}(X)| \leq O(\log (|\{x_i\}|))
\end{equation*}
\begin{proof}
First we note that $\sum n_j = |\{x_i\}|$ and, given that the set 
of matrices of size $d \times d$ is finite and so is the maximum value for $CTM(r_j)$, there exists a constant $c_d$ such that $|Adj(X)_{\{x_i\}}| CTM(r_j) < c_d$. Therefore:
\begin{equation*}
\begin{split}
BDM_{\{x_i\}}(X)-H_{\{x_i\}}(X)\\= \sum (CTM(r_j) + \log
(n_j) + \frac{n_j}{|\{x_i\}|} \log(\frac{n_j}{|\{x_i\}|})
)\\
\leq c_d + \sum (\log (n_j) +  \frac{n_j}{|\{x_i\}|}
\log (\frac{n_j}{|\{x_i\}|}) )\\
= c_d + \sum (\log (n_j) -  \frac{n_j}{|\{x_i\}|}
\log (\frac{|\{x_i\}|}{n_j}) )\\
= c_d + \frac{1}{|\{x_i\}|}\sum (|\{x_i\}|\log (n_j) - n_j
\log (\frac{|\{x_i\}|}{n_j}) ) \\
= c_d + \frac{1}{|\{x_i\}|} \sum
\log(\frac{n_j^{|\{x_i\}|+n_j}}{|\{x_i\}|^{n_j}})
\end{split}
\end{equation*}
\noindent{}Now, let's recall that the sum of $n_j$'s is bounded 
by $|\{x_i\}|$. Therefore there exists  $c'_d$ such that
\begin{align*}
\frac{1}{|\{x_i\}|} \sum
\log (\frac{n_j^{|\{x_i\}|+n_j}}{|\{x_i\}|^{n_j}})
 & \leq
\frac{c_d}{|\{x_i\}|}
\log (\frac{|\{x_i\}|^{|\{x_i\}|+c'_d|\{x_i\}|}}{|\{x_i\}|^{c'_d|\{x_i\}|}})\\
& = \frac{c_d}{|\{x_i\}|}
\log (|\{x_i\}|^{|\{x_i\}|})\\
& = c_d \log (|\{x_i\}|).
\end{align*}
\end{proof}
\end{prop}

Now, is important to note that the previous proof sets the limit in terms of the constant $c_d$, which minimum value is defined in terms of matrices for which the $CTM$ value has been computed. The smaller this number is, the tighter is the bound set by \ref{t3}. Therefore, in the worst case, this is when $CTM$ has been computed for a comparatively small number of matrices, or the larger base matrix have small algorithmic complexity, the behavior of $BDM$ is similar to entropy. In the best case, when $CTM$ is updated by any means, BDM approximates algorithmic complexity (corollary \ref{t2}). 

Furthermore, we can think on $c_d \log (|\{x_i\}|)$ as a measure of the deficit in information incurred by both, entropy and BDM, in terms of each other. Entropy is missing the number of base objects needed in order to get an approximation of the compression length of $M$, while BDM is missing the position of each base symbol. And giving more information to both measures wont necessarily yield a better approximation to $K$.

\section{Normalized BDM}
\label{norm-bdm-sec}
A normalized version of BDM is useful for applications in which a maximal value of complexity is known or desired for comparison purposes. The chief advantage of a normalized measure is that it enables a comparison among objects of different sizes, without allowing size to dominate the measure. This will be useful in comparing arrays and objects of different sizes. First, for a square array of size $n\times n$, we define:

\begin{equation}
\label{newecaeq}
Min BDM(n)_{d\times d} = \lfloor n/d \rfloor + \displaystyle\min_{x\in M_d(\{0,1\})} CTM(x)
\end{equation}

\noindent where $M_d(\{0,1\})$ is the set of binary matrices of size $d\times d$. For any $n$, $MinBDM(n)_{d\times d}$ returns the minimum value of Eq.~\eqref{newecaeq} for square matrices of size $n$, so it is the minimum BDM value for matrices with $n$ nodes. It corresponds to an adjacency matrix composed of repetitions of the least complex $d\times d$ square. It is the all-1 or all-0 entries matrix, because $0_{d,d}$ and $1_{d,d}$ are the least complex square base matrices (hence the most compressible) of size $d$.

\begin{figure}[ht!]
\centering
\scalebox{.38}{\includegraphics{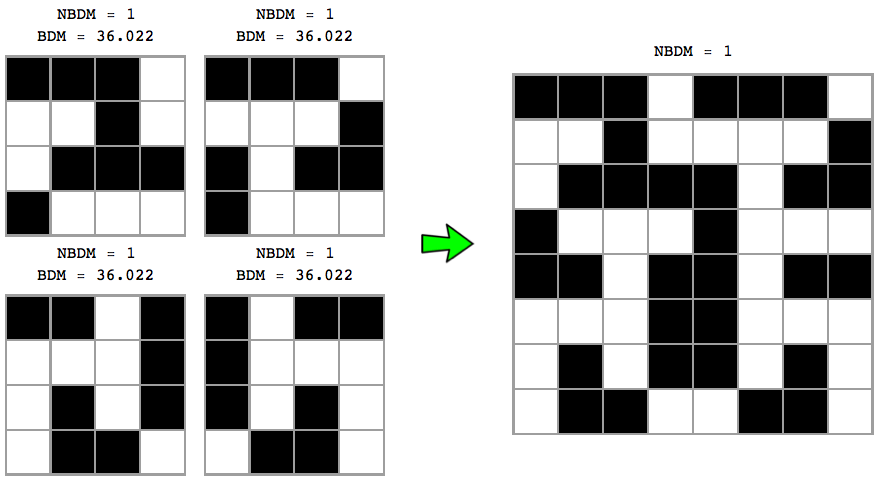}}\caption{\label{NBDM}NBDM assigns maximum value 1 to any base matrix with highest CTM or any matrix constructed out of base matrices. In this case, the 4 base matrices on the left are those with the highest CTM in the space of all base matrices of the same size, while the matrix to the left is assigned the highest value because it is built out of the maximum complexity base matrices.}
\end{figure}

Secondly, for the maximum complexity, Eq.~\eqref{newecaeq} returns the highest value when the result of dividing the adjacency matrix into the $d\times d$ base matrices contains the highest possible number of different matrices (to increase the sum of the right terms in Eq.~\eqref{newecaeq}) and the repetitions (if necessary) are homogeneously distributed along those squares (to increase the sum of the left terms in Eq.~\eqref{newecaeq}) which should be the most complex ones in $M_d(\{0,1\})$. For $n,d\in \mathbb{N}$, we define a function \[f_{n,d}: M_d(\{0,1\})\longmapsto \mathbb{N}\]
that verifies:
\begin{equation}
  \label{eq:2}
   \displaystyle\sum_{r\in M_d(\{0,1\})} f_{n,d}(r) = \lfloor
  n/d\rfloor^2  \\
\end{equation}
  \begin{equation}
  \label{eq:3}
   \displaystyle\max_{r\in M_d(\{0,1\})} f_{n,d}(r)\\ \leq\ \ 1+
  \displaystyle\min_{r\in M_d(\{0,1\})} f_{n,d}(r)   \\
 \end{equation}
 \begin{equation}  
 \label{eq:4}
  CTM(r_i) > CTM(r_j)\ \Rightarrow \ f_{n,d}(r_i)\geq
  f_{n,d}(r_j)
\end{equation}

The value $f_{n,d}(r)$ indicates the number of occurrences of $r\in
M_d(\{0,1\})$ in the decomposition into $d\times d$ squares of the
most complex square array of size $n\times n$. Condition Eq.~\eqref{eq:2}
establishes that the total number of component squares is $\lfloor
n/d\rfloor^2$. Condition Eq.~\eqref{eq:3} reduces the square repetitions 
as much as possible, so as to increase the number of differently composed squares as far as possible and distribute them homogeneously. Finally, Eq.~\eqref{eq:4} ensures that the most complex squares are the best represented. Then, we define:
\[
Max BDM(n)_{d\times d} = \hspace{-0.5cm}\sum_{
  {\begin{array}{c}
    r\in M_d(\{0,1\}),\\f_{n,d}(r)>0
  \end{array}}} \hspace{-0.5cm}
\log_2(f_{n,d}(r))+ CTM(r)
\]
Finally, the normalized BDM value of an array $X$ is:\\

\begin{definition}
Given an square matrix $X$ of size $n$,
$NBDM(X)_d$ is defined as 
\begin{equation}
\label{nbdm}
\frac{CTM(X) - Min BDM(n)_{d\times
    d}}{Max BDM(n)_{d\times d} - Min BDM(n)_{d\times d}}
\end{equation}
\end{definition}
This way we take the complexity of an array $X$ to have a normalized value which is not dependent on the size of $X$ but rather on the relative complexity of $X$ with respect to other arrays of the same size. Fig.~\ref{NBDM}, provides an example of high complexity for illustration purposes. The use of $Min BDM(n)_{d\times d}$ in the normalization is relevant. Note that the growth of $Min BDM(n)_{d\times d}$ is
 linear with $n$, and the growth of $Max BDM(n)_{d\times d}$ exponential. This means that for high complexity matrices, the result of normalizing by using just $CTM(X)/Max BDM(n)_{d\times d}$ would be similar to $NBDM(X)_d$. But it would not work for low complexity arrays, as when the complexity of $X$ is close to the minimum, the value of $CTM(X)/Max BDM(n)_{d\times d}$ drops exponentially with $n$. For example, the normalized complexity of an empty array (all 0s) would drop exponentially in size. To avoid this, Eq.~\eqref{nbdm} considers not only the maximum but also the minimum. 

Notice the heuristic character of $f_{n,d}$. It is designed to ensure a quick computation of $Max BDM(n)_{d\times d}$, and the distribution of complexities of squares of size $d\in\{3,4\}$ in $D(5,2)$ ensures that $Max BDM(n)_{d\times d}$ is actually the maximum complexity of a square matrix of size $n$, but for other distributions it could work in a different way. For example, condition~\eqref{eq:3} assumes that the complexities of the elements in $M_d(\{0,1\})$ are similar. This is the case for $d\in\{3,4\}$ in $D(5,2)$, but it may not be true for other distributions. But at any rate it offers a way of comparing the complexities of different arrays independent of their size.

\section{CTM to BDM Transition}

How BDM scales CTM remains a question, as does the rate at which BDM loses the algorithmic estimations provided by CTM. Also unknown is what the transition between CTM and CTM$+$BDM looks like, especially in the face of applications involving objects of medium size between the range of application of CTM (e.g. 10 to 20 bit strings) and larger objects (e.g. longer sequences in the hundreds of bits).

We perform a Spearman correlation analysis to test the strength of a monotonic relationship between CTM values and BDM values with various block sizes and block overlap configurations in all 12 bit strings. We also test the strength of this relationship with CTM on Shannon entropy and compression length.

\begin{figure}[ht!]
\centering
\includegraphics[width=0.95\textwidth]{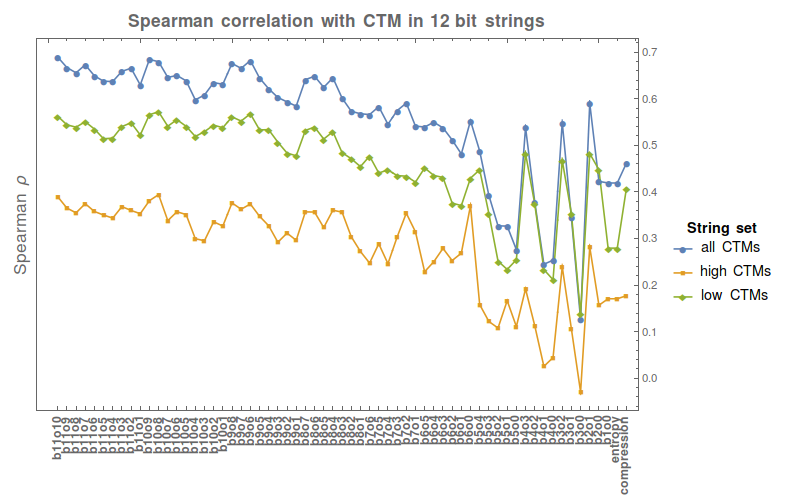}
\caption{Spearman correlation coefficients ($\rho$) between CTM and BDM of all possible block sizes and overlap lengths for 12 bit strings, compared with the correlation between CTM and Shannon entropy, and the correlation between CTM and compression length (shown at the rightmost edge of the plot) in blue. $\rho$ coefficients for the 2048 strings below and above the median CTM value are shown in green and orange, respectively. BDM block size and overlap increases to the left.  Compression length was obtained using \textit{Mathematica}'s \texttt{Compress[]} function. All values were normalized as described in \ref{norm-bdm-sec}. }   
\label{bdm-ctm}
\end{figure}

Fig.~\ref{bdm-ctm} shows the agreement between BDM and CTM for strings for which we have exact CTM values, against which BDM was tested. The results indicate an agreement between CTM and BDM in a variety of configurations, thereby justifying BDM as an extension of the range of application of CTM to longer strings (and to longer objects in general). 

In the set of all 12 bit strings, the correlation is maximal when block size = 11 and overlap = 10 (b11o10, $\rho = 0.69$); Shannon entropy has $\rho=1$ with BDM when strings are divided in blocks of size = 1 and overlap = 0 (b1o0, $\rho = 0.42 $), as is expected from what is described in~\ref{convergence-to-entropy}.

The Spearman rank test performed on the first 4096 binary strings has p-values $< 1 \times 10^{-15}$, while the Spearman rank test on the 2048 strings with CTM below the median has p-values $< 1 \times 10^{-9}$. Finally the Spearman rank test on the 2048 strings with CTM value above the median has p-values $< 1 \times 10^{-5}$, in all cases except those corresponding to b4o1, b4o0, and b3o0, where $ \rho < 0.03$ and 0.045 $\leq$ p-value $\geq 0.25$. The lower $\rho$ coefficients in above median CTM strings indicates that there is a greater difficulty in estimating the algorithmic complexity of highly irregular strings through either BDM, entropy, or compression length than in detecting their regularity. Fig.~\ref{bdm-ctm} shows that for block size $> 6$ the Spearman $\rho$ of BDM is always higher than the correlation of CTM with either Shannon entropy or compression length. Some block configurations of size $< 6$ (e.g., b2o1) also have higher $\rho$ than both Shannon entropy and compression.

While BDM approximates the descriptive power of CTM and extends it over a larger range, we prove in Subsection~\ref{boundproof} that BDM approximates Shannon entropy if base objects are no longer generated with CTM, but if CTM approximates algorithmic complexity, then BDM does.

\subsection{Smooth BDM (and `add col')}
\label{mibdm}

As an alternative method for increasing accuracy while decreasing computational cost, is the use of a weighted function as penalization parameter in BDM. Let the base matrix size be $4\times 4$. We first  partition the matrix into sub matrices of the matrix base size $4\times 4$. If the matrix size is not divisible by $4$ we (1) use a smooth BDM with full overlap boundary condition (we call this method somply `smooth' BDM) or (2) add an artificial low complexity boundary to `complete' the matrix to the next multiple of 4 and apply `smooth' (we call this approach 'add col' in future sections).

When using the BDM full overlap boundary condition, we screen the entire matrix by moving a sliding square of size 4$\times$4 over it (as it is done for `recursive BDM'). When adding artificial low complexity boundaries we only calculate non overlapping sub-matrices of size $4\times 4$ because the expanded matrix of interest is of multiple of 4. These artificial low complexity boundary are columns and rows of one symbols (zeroes or ones). We then correct the final result by subtracting the information added to the boundaries from $\log(|R|)+\log(|C|)$.

To prevent the undesired introduction of false patterns in the `completion' process (add col), we use the minimum BDM of the extended matrix for both cases (column and rows of zeroes and ones denoted by $BDM_1(X)$ and $BDM_0(X)$ respectively). 

In both cases, to distinguish the occurrence of rare and thus highly complex patterns, we assign weights to each base matrix based on the probability of seeing each pattern, denoted by $W_{i}$, where $i$ is the index of the base matrix. We thereby effectively `smooth' the transition to decide matrix similarity, unlike the previous versions of BDM which counts multiplicity of equal matrices. Thus the main difference introduced in the `smooth' version of BDM is the penalization by base matrix (statistical) similarity rather than only perfect base matrix match. 

To simplify notation, in what follows let us denote the adjacency matrix $Adj(X)$ of a matrix $M$ simply as $M$. The \textit{smooth} version of BDM is then calculated as follows:


\begin{equation}
BDM(X)=\min(BDM_0,BDM_1)
\end{equation}
\begin{equation}
BDM_f(X) =  \displaystyle\sum_{(r_i,n_i)\in{}Adj(X)_{\{x_i\}}} BDM(r_i)\times W_i+\log(n_i)
\end{equation}

\subsection{Weighted Smooth BDM with Mutual Information}

The Smooth BDM version assigns a weight to each base matrix depending on its statistical likelihood, which is equivalent to assigning a weight based on the entropy of the base matrix over the distribution of all base matrices of $4 \times 4$. An equivalent version that is computationally more expensive is the use of classical mutual information. This is arrived at by measuring the statistical similarity between base matrices precomputed by mutual information. 

Mutual information is a measure of the statistical dependence of a random variable $X$ on a random variable $Y$ in the joint distribution of $X$ and $Y$ relative to the joint distribution of $X$ and $Y$ under an assumption of independence. If $MI(X,Y) = 0$, then $X$ and $Y$ are statistically independent, but if the knowledge of $X$ fully determines $Y$, $MI(X,Y) = 1$, then $X$ and $Y$ are not independent. Because $MI$ is symmetric $MI(X,Y)=MI(Y,X)$; if $MI(X,Y) = 1$, then knowing all about $Y$ also implies knowing all about $X$. In one of its multiple versions MI of $X$ and $Y$ can be defined as:

\begin{equation}
MI(X,Y)=H(X)-H(X|Y)
\end{equation}

\noindent where $H(X)$ is the Shannon entropy of $X$ and $H(X|Y)$ the conditional Shannon entropy of $X$ given $Y$.

In this way, statistically similar base matrices are not counted as requiring 2 completely different computer programs, one for each base matrix, but rather a  slightly modified computer program producing 2 similar matrices accounting mostly for one and for the statistical difference of the other. More precisely, BDM can be defined by:

\begin{equation}
BDM(X) = \sum_{(r_i,n_i)\in{}Adj(X)_{\{x_i\}}} MIBDM(r_i) + \log n_i
\end{equation}

\noindent where MIBDM is defined by:

\begin{align}
 MIBDM(r_i)=\min\{MI(r_i,r_j) CTM(r_i) \label{eq:MIBDM}\\
 + (1-MI(r_j,r_i)) CTM(r_j), \nonumber \\
MI(r_i,r_j) CTM(r_j) \nonumber \\
+ (1-MI(r_j,r_i)) CTM(r_i) \nonumber \}
\end{align}

\noindent and where $MI(r_i,r_j)$ is a weight for each CTM value of each base matrix such that $j$ is the index of the matrix that maximizes $MI$ (or maximizes statistical similarity) over the distribution of all the base matrices such that $MI(r_i,r_j)\geq MI(r_i,r_k)$ for all $k \in \{1, \ldots, N\}$, $N=|Adj(X)_{\{x_i\}}|$. 

However, this approach requires $N \times N$ comparisons $MI(r_i,r_j)$ between all base matrices $r$ with indexes $i \in \{1, \ldots, N\}$ and $j \in \{1, \ldots, N\}$.

Notice that because $MI$ is symmetric, then $MI(r_i,r_j)=MI(r_j,r_i)$, but the $\min$ in Eq.~\ref{eq:MIBDM} is because we look for the minimum CTM value (i.e. the length of the shortest program) for the 2 cases in which one base matrix is the one helping define the statistical similarities of the other and vice versa.

\section{Testing BDM and Boundary Condition Strategies}
\label{comparison} 

A test for both CTM and BDM can be carried out using objects that have different representations or may look very different but are in fact algorithmically related. First we will prove some theorems relating to the algorithmic complexity of dual and cospectral graphs and then we will perform numerical experiments to see if CTM and BDM perform as theoretically expected. 

A dual graph of a planar graph $G$ is a graph that has a vertex corresponding to each face of $G$, and an edge joining two neighbouring faces for each edge in $G$. If $G^\prime$ is a dual graph of $G$, then $A(G^\prime)=A(G)$, making the calculation of the Kolmogorov complexity of graphs and their dual graphs interesting--because of the correlation between Kolmogorov complexity and $A(G^\prime)$, which should be the same for $A(G)$. One should also expect the estimated complexity values of graphs to be the same as those of their dual graphs, because the description length of the dual graph generating program is $O(1)$.

\textit{Cospectral graphs}, also called \textit{isospectral} graphs, are graphs that share the same graph spectrum. The set of graph eigenvalues of the adjacency matrix is called the spectrum $Spec(G)$ of the graph $G$. This cospectrality test for complexity estimations is interesting because two non-isomorphic graphs can share the same spectrum. 

We have demonstrated that isomorphic graphs have similar complexity as a function of graph automorphism group size~\cite{zenilgraph}. We have also provided definitions for the algorithmic complexity of labelled and unlabelled graphs based on the automorphism group~\cite{zenilmethodsbiology}. In the Appendix we prove several theorems and corollaries establishing the theoretical expectation that dual and cospectral graphs have similar algorithmic complexity values, and so we have a theoretical expectation of numerical tests with BDM to compare with.

\begin{table}[htpb] 
\centering
\resizebox{\columnwidth}{!}{
    \begin{tabular}{c|c|c|c|c}
       &\textbf{Non-overlapping}&\textbf{Fully overlapping}&\textbf{Smooth fully} &\textbf{Smooth add row}  \\
       &\textbf{BDM}&\textbf{recursive BDM} &\textbf{overlapping BDM} & \textbf{or column BDM} \\
        \hline
         \hline
        \textbf{Duality test} & 0.874 & 0.783 & 0.935 & 0.931 \\ 
        \hline
        \textbf{Cospectrality test} & 0.943 & 0.933 & 0.9305 & 0.931
    \end{tabular}} 
    \caption{Spearman $\rho$ values of various BDM versions tested on dual and cospectral graphs that theoretically have the same algorithmic complexity up to a (small) constant.}
    \label{tab:graph-rhos}
\end{table}

Compression lengths and BDM values in Table~\ref{tab:graph-rhos} and Fig.~\ref{dual} (Sup. Inf.) are obtained from the adjacency matrices of 113 dual graphs and 193 cospectral graphs from \emph{Mathematica}'s \texttt{GraphData[]} repository. Graphs and their dual graphs were found by BDM to have estimated algorithmic complexities close to each other. While entropy and entropy rate do not perform well in any test compared to the other measures, compression retrieves similar values for cospectral graphs as compared to BDM, but it is outperformed by BDM on the duality test. The best BDM version for duals was different from that for cospectrals. For the duality test, the smooth, fully overlapping version of BDM outperforms all others, but for cospectrality, overlapping recursive BDM outperforms all others.

In~\cite{zenilgraph}, we showed that BDM behaves in agreement with the theory with respect to the algorithmic complexity of graphs and the size of the automorphism group to which they belong. This is because the algorithmic complexity $K(G)$ of $G$ is effectively a tight upper bound on $K(Aut(G))$.

\section{Conclusion}
We have introduced a well-grounded, theoretically sound and robust measure of complexity that beautifully connects 2 of the main branches of information theory, classical and algorithmic. We have shown that the methods are scalable in various ways, including native $n$-dimensional variations of the same measure. The properties and numerical experiments are in alignment with theoretical expectations and represent the only truly different alternative and more accurate measure of algorithmic complexity currently available. We have also shown that BDM is computationally efficient, hence complementing the effective use of lossless compression algorithms for calculation of upper bounds of Kolmogorov complexity. 

There are thus three methods available today for approximating $K$ (two of which have been advanced by us, one being completely novel: BDM; and one that was known but had never been calculated before: CTM). Here they are described by their range of application:

\begin{myitemize}
\item CTM deals with all bit strings of length 1-12 (and for some 20-30 bits).
\item BDM deals with 12 bits to hundreds of bits (with a cumulative error 
that grows by the length of the strings--if not applied in conjunction with CTM). The worst case occurs when substrings share information content with other decomposed substrings and BDM just keeps adding their $K$ values individually.
\item CTM$+$BDM (deals with any string length but it is
computationally extremely expensive)
\item Lossless compression deals with no less than 100 bits and is unstable up to about 1K bits.
\end{myitemize}

While CTM cannot produce estimations of longer bitstrings, estimating the algorithmic complexity of even bitstrings can be key to many problems. Think of the challenge posed by a puzzle of 1000 pieces, if you were able to put together only 12 local pieces at a time, you would be able to put all the puzzle together even without ever looking at the whole piece and thus not even requiring to see possible non-local long-range algorithmic patterns.

Because BDM locally estimates algorithmic complexity via algorithmic probability based upon CTM, it is slightly more independent of object description than computable measures such as Shannon entropy, though in the `worst case' it behaves like Shannon entropy. We have also shown that the various flavours of BDM are extremely robust, both by calculating theoretical errors on tensors and by numerical investigation, establishing that any BDM version is fit for use in most cases. Hence the most basic and efficient one can be used without much concern as to the possible alternative methods that could have been used in its calculation, as we have exhaustively and systematically tested most, if not all, of them.

Ready to use online and offline methods and data are being released alongside this paper. An animated video is also available at \url{http://www.complexitycalculator.org/animation}

\section*{Appendix}

\subsection{Entropy and Block Entropy v CTM}

		\begin{figure}[!htp]
		\centering
\scalebox{.3}{\includegraphics{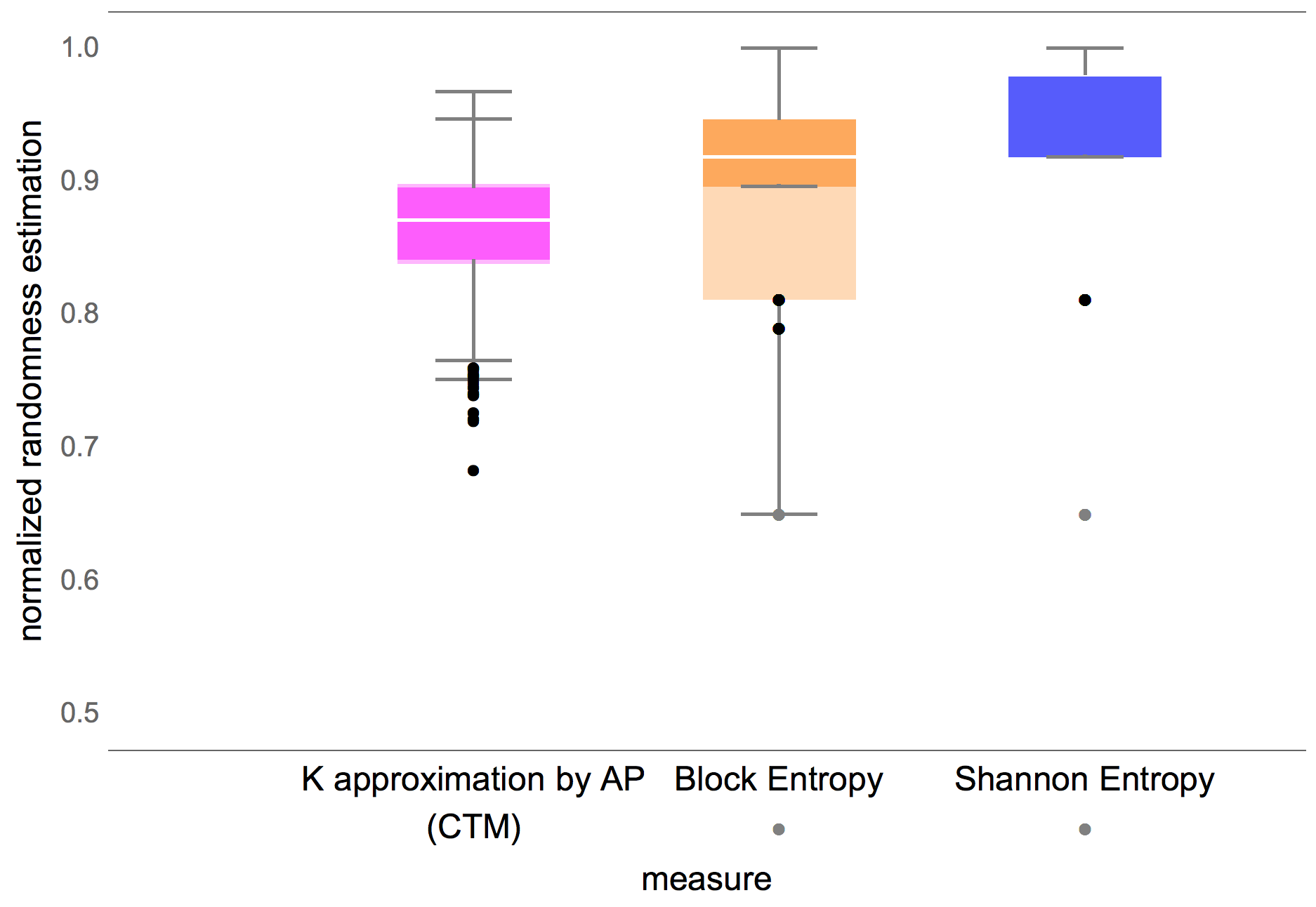}} 
\caption{\label{supfig5}\textit{The randomness of the digits of $\pi$ as measured by Shannon entropy, Block entropy and CTM. Strengthening the claim made in Fig.~\ref{piandthue}, here we show the trend of average movement of Entropy and Block Entropy towards 1 and CTM's average remaining the same but variance slightly reduced. The stronger the colour the more digits into consideration. The direction of Block entropy is the clearest, first from a sample of 100 segments of length 12 bits from the first 1000 decimal digits of $\pi$ converted to binary (light orange) followed by a second run of 1000 segments of length 12 bits from the first 1 million decimal digits of $\pi$. When running CTM over longer period of time, the invariance theorem guarantees convergence to 0.}}
		\end{figure}

\begin{center}
	\begin{table}[h]
	\caption{\label{strings}List of strings with high entropy and high Block entropy but low algorithmic randomness detected and sorted from lowest to greatest values by CTM.}
		\begin{center}
			\tabcolsep=0.11cm
			{
			\resizebox{.7\textwidth}{!}{%
			\begin{tabular}{c|c|c|c|c}
			\hline
101010010101 & 010101101010 & 101111000010 & 010000111101 & 111111000000 \\
 000000111111 & 100101011010 & 011010100101 & 101100110010 & 010011001101 \\
 111100000011 & 110000001111 & 001111110000 & 000011111100 & 111110100000 \\
 000001011111 & 111101000001 & 111100000101 & 101000001111 & 100000101111 \\
 011111010000 & 010111110000 & 000011111010 & 000010111110 & 110111000100 \\
 001000111011 & 110111000001 & 100000111011 & 011111000100 & 001000111110 \\
 110100010011 & 110010001011 & 001101110100 & 001011101100 & 111110000010 \\
 101111100000 & 010000011111 & 000001111101 & 100000111110 & 011111000001 \\
 110101000011 & 110000101011 & 001111010100 & 001010111100 & 111100101000 \\
 111010110000 & 000101001111 & 000011010111 & 111100110000 & 000011001111 \\
 110000111010 & 101000111100 & 010111000011 & 001111000101 & 111100001010 \\
 101011110000 & 010100001111 & 000011110101 & 111011000010 & 101111001000 \\
 010000110111 & 000100111101 & 111000001011 & 110100000111 & 110011100010 \\
 101110001100 & 100011001110 & 011100110001 & 010001110011 & 001100011101 \\
 001011111000 & 000111110100 & 111010000011 & 110000010111 & 001111101000 \\
 000101111100 & 110011010001 & 100010110011 & 011101001100 & 001100101110 \\
 110101001100 & 110011010100 & 001100101011 & 001010110011 & 111000110010 \\
 110010100011 & 110001010011 & 101100111000 & 010011000111 & 001110101100 \\
 001101011100 & 000111001101 & 101100001110 & 100011110010 & 011100001101 \\
 010011110001 & 111000100011 & 110001000111 & 001110111000 & 000111011100 \\
 110000011101 & 101110000011 & 010001111100 & 001111100010 & 111101010000 \\
 000010101111 & 111010001100 & 110011101000 & 001100010111 & 000101110011 \\
 111000101100 & 110010111000 & 001101000111 & 000111010011 & 111011001000 \\
 000100110111 & \text{} & \text{} & \text{} & \text{} \\
\hline
\end{tabular}%
			}
			}
			
		\end{center}
		\end{table}
\end{center}

\subsection{Duality and Cospectral Graph Proofs and Test Complement}

\begin{theorem}
Let $G^\prime$ be the dual graph of $G$. Then $K(G^\prime) \sim K(G)$.

\begin{proof}
Let $p$ denote the finite program that, for any graph $G$, replaces every edge in $G$ by a vertex and every vertex in $G$ by an edge. The resulting graph produced by $p$ is then $G^\prime$ (uniqueness), which implies that $|K(G) - K(G^\prime)| < |p|$ because we did not assume that $p$ was the shortest program. Thus, $K(G)+|p| = K(G^\prime)$ or $K(G) \sim K(G^\prime)$ up to a constant factor.
\end{proof}
\end{theorem}

Let $K(Aut(G))$ be the algorithmic complexity of the automorphism group $Aut(G)$ of the graph $G$ (i.e. all possible relabellings that preserve \textit{graph isomorphism}), that is, the length of the shortest program that generates all the graphs in $Aut(G)$.

\begin{theorem}
Let $G^\prime$ be an \textit{isomorphic graph} of $G$. Then $K(G^\prime)\sim K(G)$ for all $K(G^\prime) \in Aut(G)$, where $Aut(G)$ is the automorphism group of $G$.

The idea is that if there is a significantly shorter program $p^\prime$ for generating $G$ compared to a program $p$ generating $Aut(G)$, we can use $p^\prime$ to generate $Aut(G)$ via $G$ and a relatively short program  $c$ that tries, e.g., all permutations, and checks for isomorphism. Let's assume that there exists a program $p^\prime$ such that $||p^\prime|- |p||>c$, i.e. the difference is not bounded by any constant, and that $K(G)=|p^\prime|$. We can replace $p$ by $p^\prime + c$ to generate $Aut(G)$ such that $K(Aut(G))=p^\prime+c$, where $c$ is a constant independent of $G^\prime$ that represents the size of the shortest program that generates $Auth(G)$, given any $G$. Then we have it that $|K(Aut(G))-K(G)| < c$, which is contrary to the assumption.
\end{theorem}

\newtheorem{corollary1}{Corollary}
\begin{cor}
$K(Aut(G))<K(G^\prime)+O(1)$ for any $G^\prime \in Aut(G)$.

\begin{proof}
Let $G^\prime$ be in $Aut(G)$ such that $G \neq G^\prime$. There exists~\footnote{An algorithm (so far known to be in class \textbf{NP}) that produces all relabellings--the simplest one is brute force permutation--and can verify graph isomorphism in time class \textbf{P}~\cite{zenilgraph}).} a computer program $p$ that produces $Aut(G)$ for all $G$. With this program we can construct $Aut(G)$ from any graph $G'\in Aut(G)$ and $G'$ from $Aut(G)$ and the corresponding label $n$. Therefore $K(G') \leq |p| + K(Aut(G)) + log (n) + O(1)$ and $K(Aut(G)) <=|p| + K(G')$.
\end{proof}
\end{cor}

\begin{theorem}
If $G$ and $G^\prime$ are cospectral graphs, then $|K(G) - K(G^\prime)|<c +\log(n)$, i.e. $K(G) \sim K(G^\prime)$ up to a constant and small logarithmic term.
\begin{proof}
The strategy is similar: by brute force permutation one can produce all possible adjacency matrices after row and column permutation. Let $K(G)$ be the algorithmic complexity of $G$ and $Spec(G)$ be the spectrum of $G$. Let $p$ be the program that permutes all rows and columns and tests for cospectrality, and $|p|$ its program length. Let $Spec(G)=Spec(G^\prime)$. Then $K(G^\prime) \leq K(G)+|p|+\log(n)$ where $n$ is the size of $G$ that indicates the index of the right column and row permutation that preserves $Spec(G)$ among all graphs of size $n$.
\end{proof}
\end{theorem}

\begin{figure}[htbp!]
  \centering
Graph Duality Test\\
\includegraphics[width=\columnwidth]{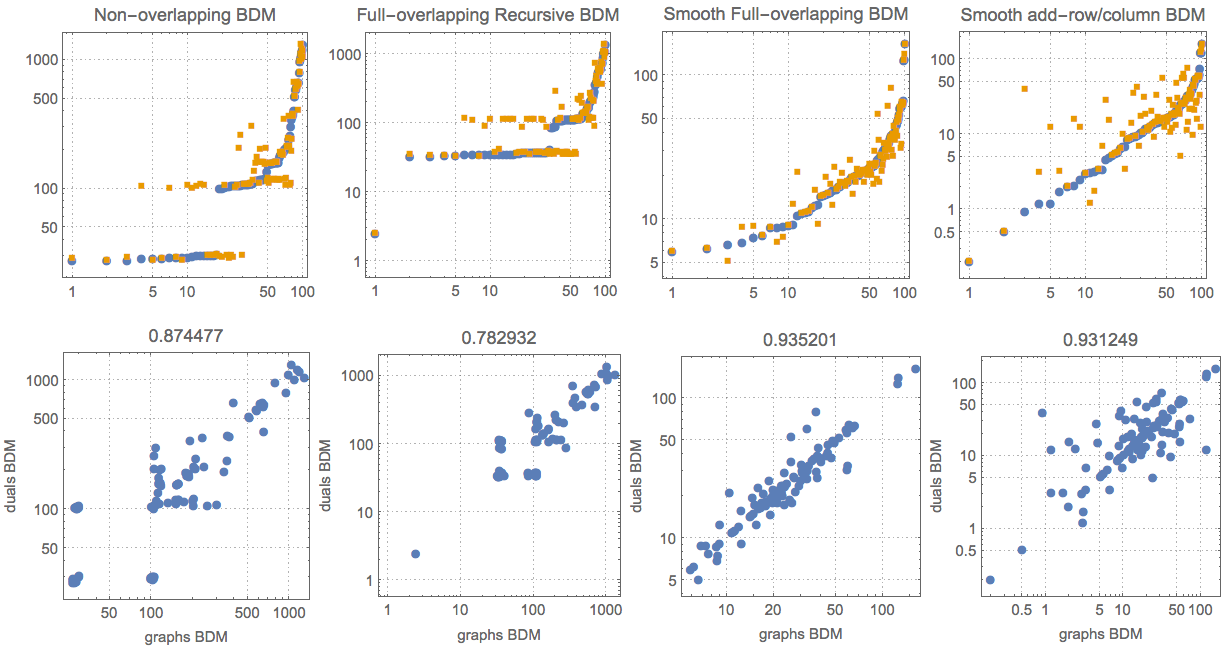}\\
Graph Cospectrality Test\\
\includegraphics[width=\columnwidth]{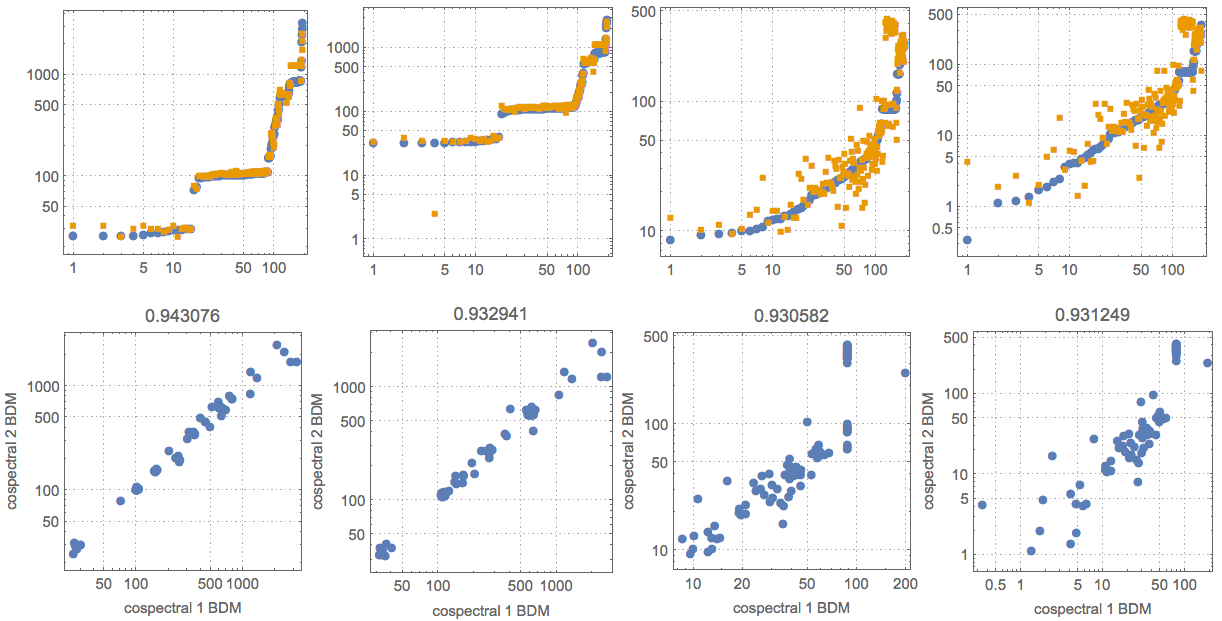}
  \caption{\label{dual} Scatterplots comparing the various BDM versions tested on dual and cospectral graphs that theoretically have the same algorithmic complexity up to a (small) constant. $x$-axis values for each top row plot are sorted by BDM for one of the dual and for the cospectral graph series. Bottom rows: on top of each corresponding scatterplot are the Spearman $\rho$ values.}
\end{figure}

\begin{figure}[htbp!]
  \centering
  Graph Duality Test\\
\includegraphics[width=\columnwidth]{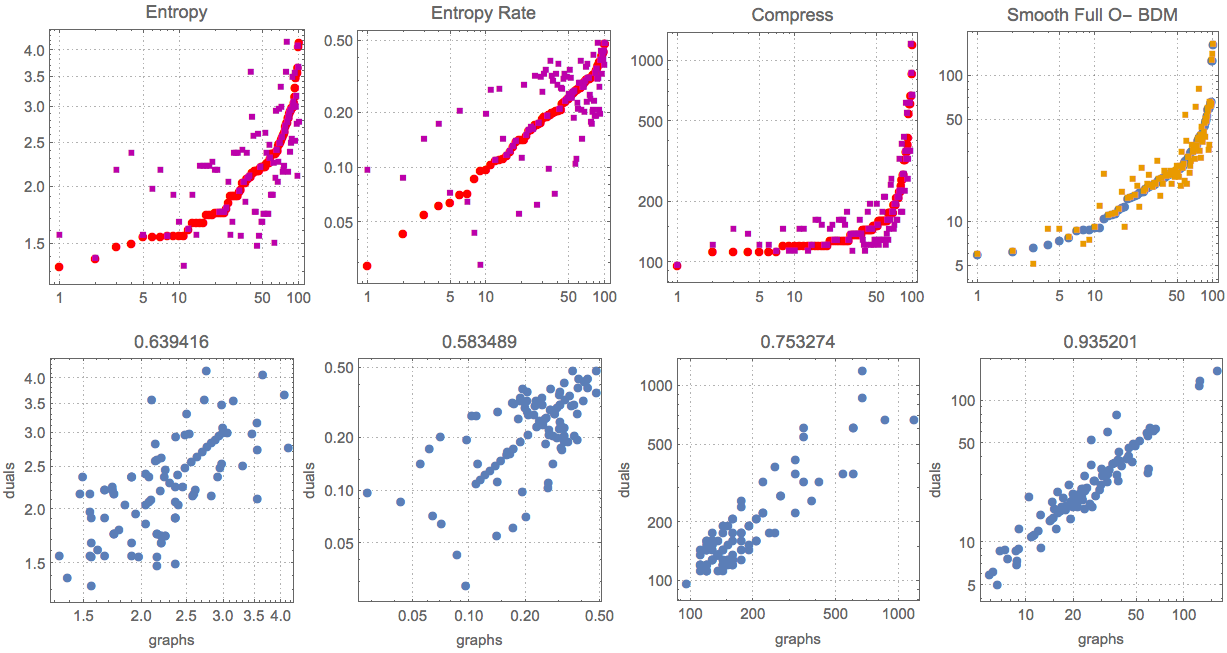}\\
Graph Cospectrality Test\\
\includegraphics[width=\columnwidth]{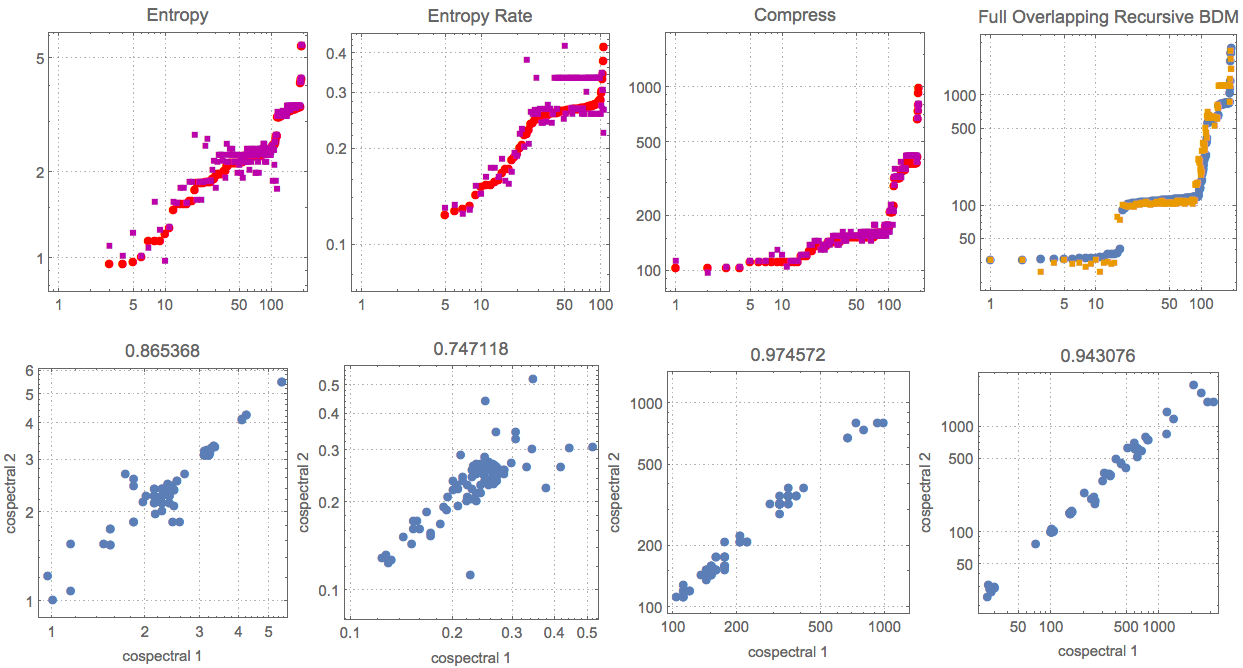}
  \caption{\label{entcomp} Scatterplots comparing other measures against the best BDM performance. $x$-axis values for each top row plot are sorted by BDM for one of the dual and for the cospectral graph series. Bottom rows: on top of each corresponding scatterplot are the Spearman $\rho$ values.}
\end{figure}

\section{The Online Algorithmic Complexity Calculator and Language Implementations}

The \textit{Online Algorithmic Complexity Calculator} (or OACC) available at~\url{http://www.complexitycalculator.com} has now been updated to incorporate 2-dimensional data, in preparation for this paper. The OACC implements the most important methods and tools explained here. Documentation is available at~\url{https://cran.r-project.org/web/packages/acss/acss.pdf}, as the OACC is an online app based on the \texttt{acss}~\cite{acss-cran} \textit{R} package. 

		\begin{figure}[!htp]
		\centering
\scalebox{.45}{\includegraphics{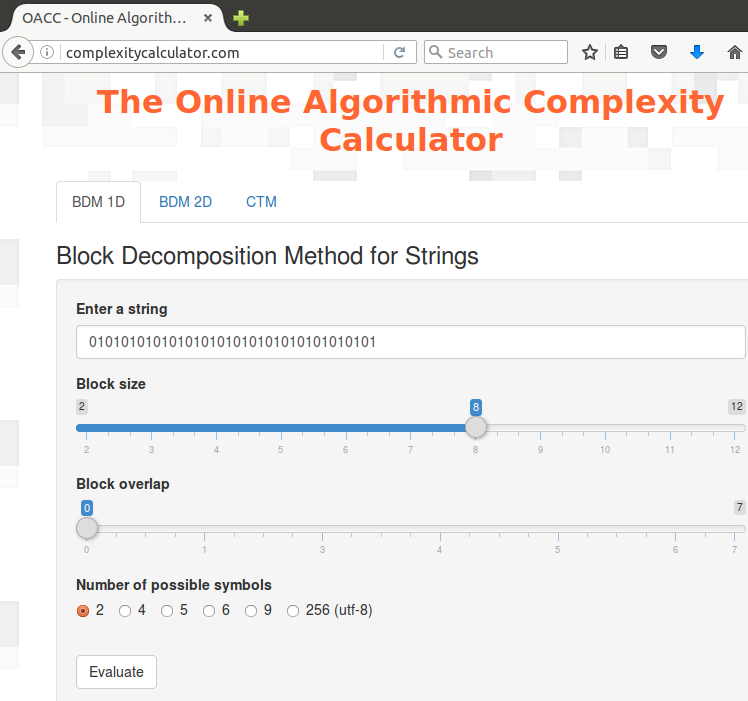}} 
\caption{\textit{The Online Algorithmic Complexity} Calculator available at \url{http://www.complexitycalculator.com}. Full code for the R Shiny web server used is available at~\cite{github-oacc}.}
		\end{figure}

We are releasing actual implementations of BDM in, \textit{C++}, \textit{Pascal}, \textit{Perl}, \textit{Python}, \textit{Mathematica} and \textit{Matlab}. Table~\ref{languages} provides an overview of the various implementations. We are also releasing the data generated by CTM which BDM requires to run. This consists in the Kolmogorov complexity evaluations of all strings up to length 11 (12 with the completion of a single one and its complement) as well as for some longer ones, and of all the square arrays of length up to 4$\times$4 as calculated by the canonical 2-dimensional Turing machine model (replacing the unidimensional tape with Rado's 2-dimensional model). The data is also released with a program (a Wolfram Language `Demonstration') available on-line at the Wolfram's Demonstrations Project website~\cite{demonstration}.

\begin{table}[htbp!]
\begin{center}
\resizebox{.9\textwidth}{!}{%
\begin{tabular}{l|c|c|c|c|c|c|c|c}
\hline
\textit{lang} & \textbf{1D n-o} & \textbf{1D}
\textbf{2D n-o} & \textbf{2D Rec} & \textbf{2D Cyc} & \textbf{2D Smo} & \textbf{2D Norm} & \textbf{1-D LD} & \textbf{addcol}\\
\hline
\textit{Online} & \ding{51} & \ding{51} & \ding{51} & $\times$ & $\times$ & $\times$ & $\times$ & $\times$ \\
\textit{WL} & \ding{51} & \ding{51} & \ding{51} & \ding{51} & \ding{51} & $\times$ & \ding{51} & $\times$\\
\textit{R} & \ding{51} & \ding{51} & \ding{51} & \ding{51} & $\times$ & $\times$ & $\times$ & $\times$\\
\textit{Matlab} & \ding{51} & $\times$ & \ding{51} & \ding{51} & $\times$ & \ding{51} & $\times$ & \ding{51}  \\
\textit{Haskell} & \ding{51} & \ding{51} & \ding{51} & \ding{51} & $\times$ & $\times$ & $\times$ & $\times$ \\
\textit{Perl} & $\times$ & $\times$ & \ding{51} & $\times$ & $\times$ & $\times$ & $\times$  & $\times$\\
\textit{Python} & $\times$ & $\times$ & \ding{51} & $\times$ & $\times$ & $\times$ & $\times$ & $\times$ \\
\textit{Pascal} & $\times$ & $\times$ & \ding{51} & $\times$ & $\times$ & $\times$ & $\times$ & $\times$ \\
\textit{C++} & $\times$ & $\times$ & \ding{51} & $\times$ & $\times$ & $\times$ & $\times$ & $\times$\\
\hline
\end{tabular}%
}
\end{center}
\caption{\label{languages}Computer programs in different languages implementing various BDM versions. We have shown that all implementations agree with each other in various degrees, with the only differences having to do with under- or over-estimated values and time complexity and scalability properties. They are extremely robust, thereby establishing that the use of the most basic versions (1D n-o, 2D n-o) are justified in most cases. `WL' stands for Wolfram Language, the language behind e.g. the \textit{Mathematica} platform, `Online' for the online calculator, `Cyc' for Cyclic, `Norm' stands for normalized, `Rec' for recursive, `Smo' for `Smooth', `N-o' for Nonoverlapping and `addcol' for the method that adds rows and columns of lowest complexity to the borders up to the base string/array/tensor size. If not stated then it supports overlapping. All programs are available at \url{https://www.algorithmicdynamics.net/software.html}}
\end{table}

\end{document}